\documentclass[12pt]{article}
\pdfoutput=1
\usepackage{amsmath}
\usepackage{amssymb}
\usepackage{graphicx}
\usepackage{subcaption}
\usepackage{url}

\usepackage[sort&compress, numbers, merge]{natbib}

\usepackage[colorlinks=true, linkcolor=blue, citecolor=blue,
urlcolor=black]{hyperref}

\begin{document}
\baselineskip 0.6cm

\def\simgt{\mathrel{\lower2.5pt\vbox{\lineskip=0pt\baselineskip=0pt
           \hbox{$>$}\hbox{$\sim$}}}}
\def\simlt{\mathrel{\lower2.5pt\vbox{\lineskip=0pt\baselineskip=0pt
           \hbox{$<$}\hbox{$\sim$}}}}
\def\simprop{\mathrel{\lower3.0pt\vbox{\lineskip=1.0pt\baselineskip=0pt
             \hbox{$\propto$}\hbox{$\sim$}}}}
\def\tr{\mathop{\rm tr}}

\begin{titlepage}

\begin{flushright}
DESY 15-045 \\
FTPI-MINN-15/15 \\
IPMU15-0037
\end{flushright}

\vskip 1.1cm

\begin{center}

{\Large \bf 
Probing Bino-Gluino Coannihilation at the LHC
}

\vskip 1.2cm

Natsumi Nagata${}^{1}$,
Hidetoshi Otono${}^{2}$, 
and
Satoshi Shirai${}^{3}$
\vskip 0.5cm

{\it
$^1$ William I. Fine Theoretical Physics Institute, School of
Physics and Astronomy, \\ 
University of Minnesota, Minneapolis, MN 55455, USA,\\ 
and 
Kavli Institute for the Physics and Mathematics of the Universe (WPI),
 \\The University of Tokyo Institutes for Advanced Study, \\ The University
 of Tokyo, Kashiwa
 277-8583, Japan\\ [5pt]
$^2$ Research Center for Advanced Particle Physics, Kyushu University, Fukuoka 812-8581, Japan\\
$^3$ {Deutsches Elektronen-Synchrotron (DESY), 22607 Hamburg, Germany}
}

\vskip 1.0cm

\abstract{
 It has been widely known that bino-like dark matter in the
 supersymmetric (SUSY) theories in general suffers from over-production. The
 situation can be drastically improved if gluinos have a mass slightly
 heavier than the bino dark matter as they reduce the dark matter
 abundance through coannihilation. In this work, we consider
 such a bino-gluino coannihilation scenario in high-scale SUSY models,
 which can be actually realized when the squark-mass scale is less than
 100--1000 TeV. We study the prospects for exploring this bino-gluino
 coannihilation scenario at the LHC. We show that the searches for
 long-lived colored particles with displaced vertices or large energy
 loss offer a strong tool to test this scenario in collider experiments. 
}

\end{center}
\end{titlepage}

%%%%%%%%%%%%%%%%%%%%%%%%%%%%%%%%%%%%%%%%%%%%%%%%%%%%
\section{Introduction}
\label{sec:intro}
%%%%%%%%%%%%%%%%%%%%%%%%%%%%%%%%%%%%%%%%%%%%%%%%%%%%

The first stage of the LHC running has pointed a possible direction for
the actual realization of the supersymmetric (SUSY) Standard Model
(SM). First and foremost, the observed SM-like Higgs boson
\cite{Aad:2012tfa,*Chatrchyan:2012ufa} with a mass of
about $125$~GeV \cite{Aad:2015zhl} implies that the mass scale
of SUSY particles is higher than the electroweak scale; the
radiative corrections by stops easily lift up the Higgs mass from the
tree-level value predicted to be less than the $Z$-boson mass in the minimal
SUSY SM \cite{Inoue:1982ej, *Flores:1982pr}, if the stop masses
are far above the electroweak scale
\cite{Okada:1990vk,*Okada:1990gg,*Ellis:1990nz, *Haber:1990aw,
*Ellis:1991zd,Giudice:2011cg}. This is in fact consistent with
lack of any evidence in the SUSY searches so far \cite{Aad:2014wea,
Chatrchyan:2014lfa}. A relatively high
SUSY breaking scale  
offers further advantages for SUSY SMs. For instance, heavy masses of SUSY
particles suppress the flavor changing neutral current processes as well
as the electric dipole moments of the SM particles \cite{Gabbiani:1996hi,
Moroi:2013sfa, *McKeen:2013dma, *Altmannshofer:2013lfa, *Fuyuto:2013gla,
*Tanimoto:2014eva,*Tanimoto:2015ota},
which are stringently constrained by the low-energy precision experiments. 
Moreover, such heavy SUSY particles reduce the proton decay rate via
the color-triplet Higgs exchange \cite{Liu:2013ula, *Hisano:2013exa,
*Dine:2013nga, *Nagata:2013sba, *Hall:2014vga, *Evans:2015bxa} and make
the simplest version of the SUSY grand unification model
\cite{Sakai:1981gr, *Dimopoulos:1981zb} viable. In cosmology, the
gravitino problem is evaded when the gravitino mass is high enough
\cite{Kawasaki:2008qe}. These attractive points stimulate quite a few
studies of high-scale SUSY models \cite{Wells:2003tf,*Wells:2004di,
ArkaniHamed:2004fb,*Giudice:2004tc,*ArkaniHamed:2004yi,*ArkaniHamed:2005yv,
Hall:2011jd, Hall:2012zp, Ibe:2006de, *Ibe:2011aa, *Ibe:2012hu,
Arvanitaki:2012ps, ArkaniHamed:2012gw, Evans:2013lpa, *Evans:2013dza,
*Evans:2013uza,*Evans:2014hda, Nomura:2014asa}.

An order parameter of SUSY breaking is the gravitino mass $m_{3/2}$. If
the SUSY breaking effects are transmitted to the visible sector via the
gravitational interactions (or other interactions suppressed by 
some high-scale cutoff  such as the Planck scale), 
then the soft SUSY-breaking scalar masses are induced with
their size being ${\cal O}(m_{3/2})$. In this case, the scalar SUSY
particles typically have masses of the order of $m_{3/2}$; from now
on, we express the typical masses of these scalar particles by
$\widetilde{m}\sim m_{3/2}$. The masses of the fermionic SUSY
particles (gauginos and Higgsinos) are, on the other hand, dependent on
models, since their mass terms can be suppressed if there exist
additional symmetries. For example, the gaugino masses become much
smaller than the gravitino mass if the SUSY breaking fields are
charged under some symmetry. In this case, these masses are generated by
quantum effects, such as anomaly mediation contribution \cite{Giudice:1998xp,
Randall:1998uk} and threshold corrections at the SUSY breaking scale
\cite{Pierce:1996zz, Giudice:1998xp}. They are also affected by the
presence of extra particles \cite{Pomarol:1999ie,*Nelson:2002sa,
*Hsieh:2006ig,*Gupta:2012gu, *Nakayama:2013uta,*Harigaya:2013asa,
*Evans:2014xpa}. Moreover, the Higgsino mass can be suppressed by,
\textit{e.g.}, the Peccei-Quinn symmetry \cite{Peccei:1977hh} and be
much lighter than $m_{3/2}$ and $\widetilde{m}$. See for instance
Refs.~\cite{Bae:2014yta, Evans:2014pxa} for a concrete realization of
light Higgsinos.

Possible deviation of the masses of the fermionic SUSY partners from
$m_{3/2}$ and $\widetilde{m}$ gives additional benefits to SUSY
SMs. Firstly, if gauginos lie around ${\cal O}(1)$~TeV, gauge coupling
unification is realized with great precision \cite{Hisano:2013cqa} even
when the scalar mass scale $\widetilde{m}$ is much higher than the
electroweak scale. Secondly, the neutral components of these fermions,
the neutral bino, wino, and Higgsino, can be a candidate for dark matter (DM) 
in the Universe. Among them, the neutral wino is one of the 
most promising candidates since the anomaly mediation mechanism naturally
makes the wino be the lightest SUSY particle (LSP). Its thermal relic
abundance actually explains the observed DM density if the wino mass is
around 3~TeV \cite{Hisano:2006nn}. Currently the mass of the wino LSP
$M_{\tilde{W}}$ is restricted by the direct search at the LHC as
$M_{\tilde{W}}> 270$~GeV \cite{Aad:2013yna}. The wino DM scenario is
also being constrained by the indirect DM searches using gamma rays
\cite{Cohen:2013ama,*Fan:2013faa,*Hryczuk:2014hpa,
Bhattacherjee:2014dya}. These experiments, as well as the DM direct
detection experiments \cite{Hisano:2010fy,*Hisano:2010ct,*Hisano:2011cs,
*Hisano:2012wm,*Hisano:2015rsa}, can probe this scenario in
future. Higgsino DM with a mass of $\sim 1$~TeV can also account for the
observed DM density \cite{Cirelli:2007xd}. For the recent study of the
phenomenology and future prospects for this Higgsino DM scenario, see
Ref.~\cite{Nagata:2014wma} and references therein. 

The last possibility is bino DM. 
If the scalar SUSY particles and Higgsino are significantly heavy,
bino DM is usually over-produced as the interactions of bino with the SM sector tend to be suppressed.
To avoid the over-production and get correct dark matter abundance, 
we need some exceptional mechanism to reduce the bino abundance, such as coannihilation 
and Higgs funnel \cite{Griest:1990kh}.
If the Higgsino mass is heavier than $\mathcal{O}(10)$ TeV, the remaining possibility is the coannihilation.
In this case, its thermal relic agrees to the observed value if there exist some
particles degenerate with the bino DM in mass.
In fact, as 
shown in Refs.~\cite{Baer:2005jq, Profumo:2004wk, Feldman:2009zc,
deSimone:2014pda, Harigaya:2014dwa, Ellis:2015vaa}, bino DM can
explain the correct DM density if wino or gluino has a mass slightly
above the bino mass.
After all, there are various options for
DM candidates in the high-scale SUSY scenario, and therefore it is quite
important to experimentally examine each possibility.

Among the possibilities mentioned above, the collider testability of the
bino-gluino coannihilation is expected to be the most promising since
this case requires light gluinos. As we shall see below, we expect an
${\cal O}(1)$~TeV gluino mass in this case, which can be within the
reach of the LHC. This could be compared to
other DM scenarios in high-scale SUSY models; for instance, if the
gaugino masses follow the spectrum predicted by the anomaly mediation,
wino is the LSP and it becomes the main component of DM if it has a mass
of 3~TeV, as mentioned above. In this case, the gluino mass is predicted
to be ${\cal O}(10)$~TeV, which is of course far above the possible
reach of the LHC. In this sense, it could be much easier to look for
gluinos in the bino-gluino scenario than other cases. This naive
expectation, however, turns out to be questionable. The bino-gluino
coannihilation scenario requires that the mass difference between bino
and gluino, $\Delta M$, be $\Delta M \lesssim 100$~GeV. Such small mass
difference results in soft jet emissions, which make it extremely
challenging to detect the signal of gluino production. For this reason,
previous studies have concluded that it is difficult to probe this bino-gluino
coannihilation scenario at the LHC if the DM mass is heavier than 1~TeV
\cite{Harigaya:2014dwa,Bhattacherjee:2013wna,Low:2014cba}.

In this work, we show that this small mass difference actually helps us
to probe the bino-gluino coannihilation. When $\Delta M \lesssim
100$~GeV and the sfermion masses are much heavier than the gaugino masses, the
lifetime of gluinos $\tau_{\tilde{g}}$ can be long enough to distinguish
its decay signal from that of prompt decay. As will be shown below, we
expect its decay length to be $c\tau_{\tilde{g}}\gtrsim {\cal
O}(1)$~mm when the sfermion masses are ${\cal
O}(100)$~TeV. A decay length of this order is in fact the main target of
searches for long-lived colored particles with displaced vertices (DVs)
\cite{Aad:2015rba} and large energy loss
\cite{ATLAS-CONF-2015-013}. We will find that this search technique 
indeed gives a stringent limit on the bino-gluino coannihilation region,
and probe wide range of the parameter space in future experiments.

This paper is organized as follows. In the next section, we consider the
bino-gluino coannihilation scenario and show the parameter region which
accomplishes the correct DM density. The lifetime of gluino predicted
in this parameter region is given in
Sec.~\ref{sec:gluinolifetime}. Then, in Sec.~\ref{sec:results}, we
discuss the strategy of the long-lived gluino searches at the LHC, and
present the current constraint and future prospects for the bino-gluino
coannihilation scenario. Finally, Sec.~\ref{sec:conclusion} is devoted
to conclusion and discussion.

%%%%%%%%%%%%%%%%%%%%%%%%%%%%%%%%%%%%%%%%%%%%%%
\section{Bino-gluino coannihilation}
%%%%%%%%%%%%%%%%%%%%%%%%%%%%%%%%%%%%%%%%%%%%%%%%

To begin with, let us discuss the bino-gluino coannihilation scenario
\cite{Profumo:2004wk, Feldman:2009zc, deSimone:2014pda,
Harigaya:2014dwa, Ellis:2015vaa} to clarify the target parameter space
we consider in the following analysis. Throughout this paper, bino is
assumed to be the LSP and be the DM in the Universe. We consider the
case where the bino-gluino coannihilation is effective so that the
thermal relic abundance of the bino LSP is consistent with the observed
DM density $\Omega_{\text{DM}} h^2 = 0.12$. Thus, bino and gluino should be
degenerate in mass, \textit{i.e.}, $\Delta M \equiv
M_{\tilde{g}}-M_{\tilde{B}}\lesssim 100$~GeV, with $M_{\tilde{g}}$ and
$M_{\tilde{B}}$ being the gluino and bino masses, respectively. We
further assume that the typical mass of scalar SUSY particles,
$\widetilde{m}$, as well as the Higgsino mass $M_{\tilde{H}}$, is as high as the
gravitino mass $m_{3/2}$. This setup is realized with a generic
K\"{a}hler potential. The gaugino masses are supposed to be
suppressed by a loop factor compared with $m_{3/2}$, which occurs when
the SUSY breaking superfields are non-singlet. Namely, we require
$M_{\tilde{B}} \sim M_{\tilde{g}} \ll \widetilde{m}\sim M_{\tilde{H}}
\sim m_{3/2}$ in what follows. 
Moreover, we assume the wino is heavy enough not to contribute to the coannihilation process.
It turns out that such a mass spectrum can be in fact 
realized in the high-scale SUSY models \cite{Harigaya:2013asa,
Harigaya:2014dwa, Evans:2014xpa}. We will see below that the scalar mass
scale $\widetilde{m}$ gives the significant effects on the determination
of the bino DM abundance.\footnote{While completing this manuscript, we
received Ref.~\cite{Ellis:2015vaa}, which also discusses the squark mass
effects in the gluino coannihilation scenario.}

The relevant annihilation processes to the computation of the thermal
relic abundance are the self-annihilation and coannihilation of bino and
gluinos. Among them, gluino self-annihilation is the most effective
because of the strong interaction, and this plays the dominant role in
the determination of the bino relic abundance. The bino
self-annihilation and bino-gluino annihilation are much smaller than the
gluino self-annihilation, since these cross sections are suppressed by
heavy Higgsino and sfermion masses. Hence, these annihilation
processes scarcely affect the following calculation. 

An important caveat here is that the bino-gluino coannihilation does not
work efficiently without chemical equilibrium between bino and gluinos
\cite{Ellis:2015vaa, Farrar:1995pz, *Chung:1997rq}. Therefore we should
require that the transition rate between them should be fast 
enough compared to the Hubble expansion rate. The transition rate is,
however, again suppressed by heavy squark masses. Thus, we obtain an
upper bound on $\widetilde{m}$ by imposing the above condition. 
The transition rate of bino into gluino via quark scattering,
$\Gamma(\widetilde{B}q \to \widetilde{g}q)$, is estimated by the product
of the corresponding scattering cross section, $\sigma (\widetilde{B}q
\to \widetilde{g}q)$, and the number density of initial state quarks,
$n_q$. The former is approximately given by $\sigma (\widetilde{B}q \to
\widetilde{g}q) \sim T^2/ \widetilde{m}^4$ with $T$ being the
temperature of the Universe, while the latter is $n_q \sim T^3$ since
quarks are relativistic when the transition process is
active. Consequently, the transition rate is given by 
\begin{align}
\Gamma(\widetilde{B}q \to \widetilde{g}q ) \sim
 \frac{T^5}{\widetilde{m}^4} ~.
\end{align}
On the other hand, the Hubble rate $H$ goes like $H\sim
T^2/M_{\text{Pl}}$ with $M_{\text{Pl}}$ the Planck scale in the
radiation dominated epoch. 
In order to sufficiently reduce the bino density through coannihilation,
the condition $\Gamma(\widetilde{B}q \to \widetilde{g}q )  \gg H$ should
be satisfied until the bino DM decouples from thermal bath at the freeze-out
temperature $T_f\sim M_{\tilde{B}}/20$. This reads
\begin{equation}
\frac{\widetilde{m}^4}{M_{\text{Pl}}} \lesssim 
\biggl(\frac{M_{\tilde{B}}}{x_f}\biggr)^3 ~,
\end{equation}
which then gives an upper bound on the scalar mass scale
$\widetilde{m}$. Here
$x_f \equiv M_{\tilde{B}}/T_f \sim 20$. Numerically, we have
\begin{equation}
 \widetilde{m} \lesssim 250 \times
  \biggl(\frac{M_{\tilde{B}}}{1~\text{TeV}}\biggr)^{\frac{3}{4}}~~
  \text{TeV} ~.
\end{equation}
We find that when the DM mass is ${\cal O}(1)$~TeV the upper bound on
the scalar mass scale lies around ${\cal O}(10^{(2-3)})$~TeV; indeed,
many high-scale SUSY models \cite{Wells:2003tf,*Wells:2004di,
ArkaniHamed:2004fb,*Giudice:2004tc,*ArkaniHamed:2004yi,*ArkaniHamed:2005yv,
Hall:2011jd, Hall:2012zp, Ibe:2006de, *Ibe:2011aa, *Ibe:2012hu,
Arvanitaki:2012ps, ArkaniHamed:2012gw, Evans:2013lpa, *Evans:2013dza,
*Evans:2013uza,*Evans:2014hda, Nomura:2014asa} predict the SUSY breaking
scale to be this order, with which the 125~GeV Higgs mass is naturally
accounted for.  
Therefore, it is quite important to take into
account the constraint on $\widetilde{m}$ when we discuss the
bino-gluino annihilation in the high-scale SUSY scenario. 

%%%%%%%%%%%%%%%%%%%%%%%%%%%%%%%%%%%%%%%%%%%%%%%%%%%%%%%%%%%%%%%
\begin{figure}[t]
\centering
\includegraphics[clip, width = 0.6 \textwidth]{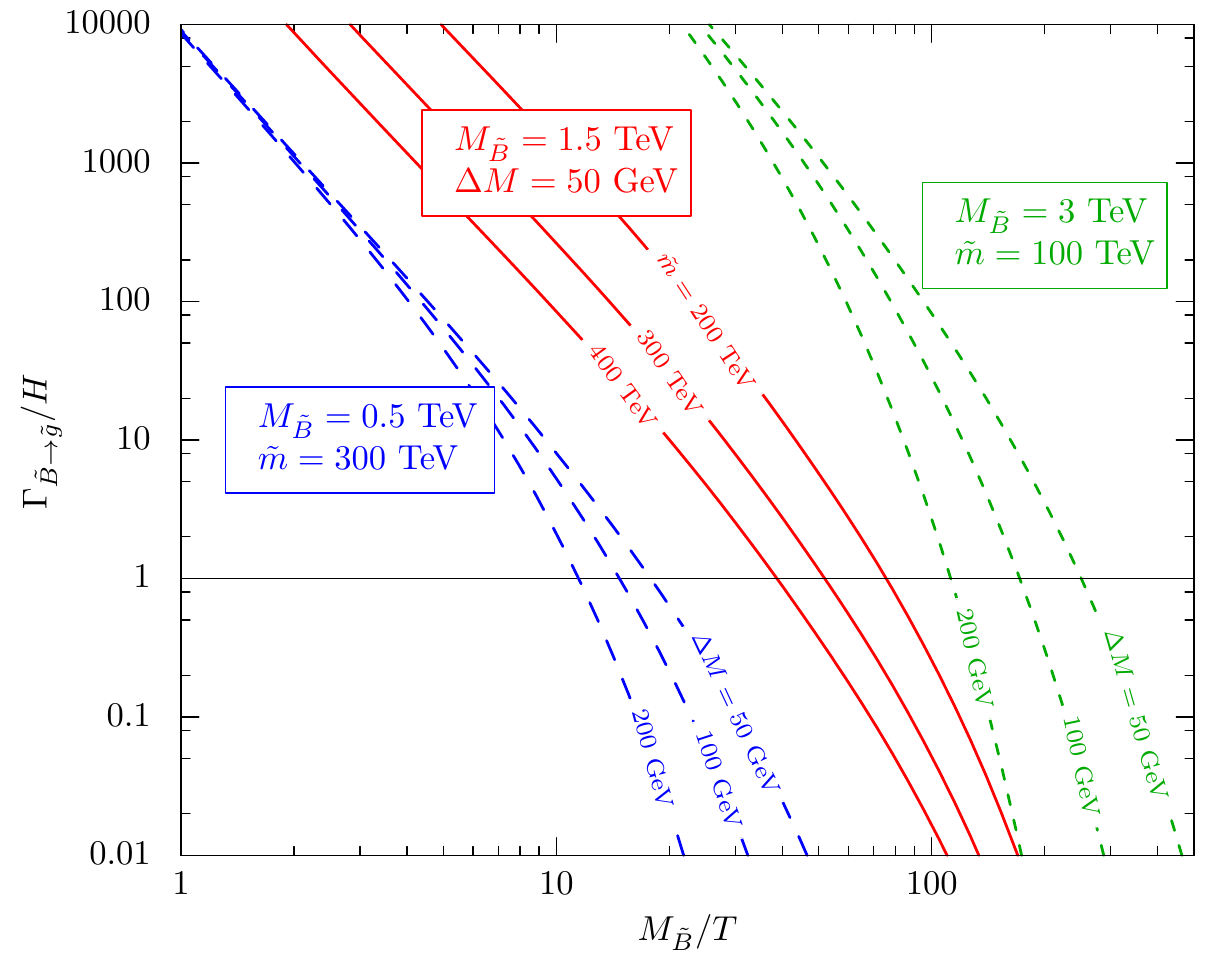}
\caption{Ratio of the bino-gluino conversion rate to the Hubble rate as
 functions of $M_{\tilde{B}}/T$.
We set $M_{\tilde B} = 1.5$~TeV, $\Delta M = 50$~GeV and $\widetilde
 m=(200,300,400)$ TeV in the red solid lines; $M_{\tilde B} = 0.5$~TeV,
 $\widetilde m = 300$ TeV and $\Delta M = (50,100,200)$ GeV in blue and
 dashed lines and $m_{\tilde  B} = 3$ TeV, $\tilde m = 100$ TeV and
 $\Delta M = (50,100,200)$ GeV in green and dotted lines. }
\label{fig:rate}
\end{figure}
%%%%%%%%%%%%%%%%%%%%%%%%%%%%%%%%%%%%%%%%%%%%%%%%%%%%%%%%%%%%%%

To make the above discussion more accurately, we perform the numerical
computation by solving the Boltzmann equation to obtain the bino-gluino
conversion rate and the resultant relic abundance. First, in
Fig.~\ref{fig:rate}, we show the ratio of the bino-gluino conversion rate
$\Gamma_{\tilde B \to \tilde g}$ with respect to the Hubble rate $H$ as
functions of $M_{\tilde{B}}/T$. Here, we set $M_{\tilde B} = 1.5$~TeV,
$\Delta M = 50$~GeV and $\widetilde m=(200,300,400)$~TeV in the red
solid lines; $M_{\tilde  B} = 0.5$~TeV, $\widetilde m = 300$~TeV and
$\Delta M = (50,100,200)$~GeV in the blue dashed lines; $M_{\tilde B} =
3$~TeV, $\widetilde m = 100$~TeV and $\Delta M = (50,100,200)$~GeV in
the green dotted lines. All of the squark masses are assumed to be equal
to the universal mass $\widetilde m$. When we evaluate the transition
cross sections and (inverse) decay rate of gluino and bino, we use the
effective theoretical approach to properly deal with sizable quantum
corrections resulting from large difference between the gluino and
squark mass scales; we first integrate out squarks to obtain
a set of dimension-six operators which involve quarks, bino and
gluino, and then evolve these operators down to the gluino mass scale by
using the 
renormalization group equations, which results in a several tens percent
enhancement of the transition rate, compared to the tree level
calculation \cite{Gambino:2005eh,Sato:2012xf,Sato:2013bta}. 
The loop-induced dimension-five dipole operator (gluon-bino-gluino) is
found to be quite suppressed and thus its contribution is negligible in
the present analysis. In addition, we include the so-called Sommerfeld
effects \cite{Hisano:2003ec, *Hisano:2004ds} on the gluino
annihilation. On top of that, $p$-wave contribution, finite-temperature
effects, the scale dependence of the strong coupling constant in the QCD
potential \cite{deSimone:2014pda}, possible ambiguity in the initial
state color arrangement\footnote{In our computation, we assume that the
initial state gluinos have a definite color configuration, not
thermal averaged one. } due to thermal effects \cite{Ibarra:2015nca},
and the bound-state effects on a pair
of gluinos \cite{Ellis:2015vaa} may change the results by a factor of
${\cal O}(10)$\%. 
The above figure shows that the conversion rate
decreases as $\widetilde{m}$ or $\Delta M$ is taken to be larger. In
particular, if the squark mass scale $\widetilde{m}$ is several hundred
of TeV with the DM mass being a relatively small, then the condition
$\Gamma_{\tilde B \to \tilde g} \gg H$ does not hold any more when the
DM abundance freezes out.

%%%%%%%%%%%%%%%%%%%%%%%%%%%%%%%%%%%%%%%%%%%%%%%%%%%%%%%%%%%%%%%
\begin{figure}[t]
\centering
\includegraphics[clip, width = 0.6 \textwidth]{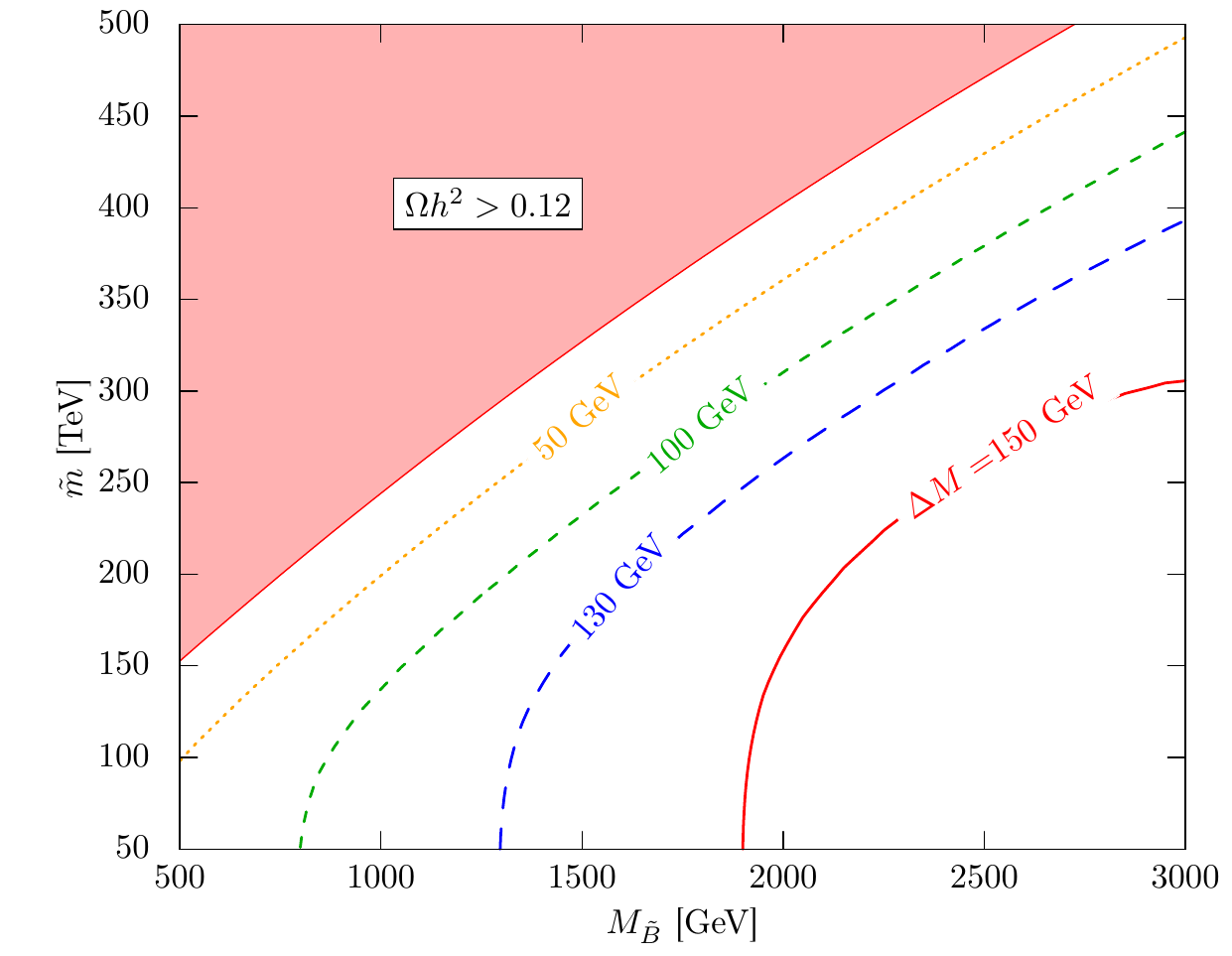}
\caption{Contour for the mass difference $\Delta M$ which makes the
 thermal relic abundance of bino DM equal to the observed DM density
 $\Omega_{\rm DM}h^2 = 0.12$. In the red shaded region the bino DM is
 overproduced due to failure of bino-gluino coannihilation. }
\label{fig:sq}
\end{figure}
%%%%%%%%%%%%%%%%%%%%%%%%%%%%%%%%%%%%%%%%%%%%%%%%%%%%%%%%%%%%%%

In Fig.~\ref{fig:sq}, we plot on the $M_{\tilde B} - \widetilde m$ plane the
mass difference $\Delta M$ with which the thermal relic abundance of
bino DM explains the observed DM density $\Omega_{\rm DM}h^2 = 0.12$.
In the red shaded region, the squark mass is too heavy for the
coannihilation process to work well and therefore the DM is
overproduced. We will discuss how to probe the parameter space shown in
Fig.~\ref{fig:sq} at the LHC in the subsequent section.

%%%%%%%%%%%%%%%%%%%%%%%%%%%%%%%%%%%%%%%%%%%%%%%%%%%%%%%%%%
\section{Gluino lifetime}
\label{sec:gluinolifetime}
%%%%%%%%%%%%%%%%%%%%%%%%%%%%%%%%%%%%%%%%%%%%%%%%%%%%%%%%%%

Next, we study the lifetime of gluino, which plays a crucial role in the
discussion of the testability of the bino-gluino coannihilation
scenario at the LHC in the following section.
As mentioned in the Introduction, in this scenario,
a relatively light gluino mass is expected. Thus, the gluino pair
production is suitable target for the hadron collider experiments like
the LHC in this case. After the pair production, a gluino decays 
into a bino, a quark, and an anti-quark through the squark-exchange processes
\cite{Toharia:2005gm,Gambino:2005eh,Sato:2012xf,Sato:2013bta}. When the
gluino is degenerate with the bino in mass, which is required in the
bino-gluino coannihilation scenario, the decay length of the gluino, $c
\tau_{\tilde{g}}$, is approximately given as follows:
\begin{align}
c\tau_{\tilde g} = {\cal O}(1)\times \left(\frac{\Delta
 M}{100~\mathrm{GeV}}\right)^{-5} \left(\frac{\widetilde{m}}
{100~\mathrm{TeV}}\right)^4 ~\text{cm} ~.
\label{eq:ctauglu}
\end{align}
From this equation, we see that the decay length gets longer as the
mass difference $\Delta M$ is taken to be smaller or the scalar mass scale
$\widetilde{m}$ is set to be larger. Therefore, we expect a relatively
long decay length when the bino-gluino coannihilation is achieved in the
high-scale SUSY scenario.

%%%%%%%%%%%%%%%%%%%%%%%%%%%%%%%%%%%%%%%%%%%%%%%%%%%%%%%%%%%%%%%%%
\begin{figure}[t]
\centering
\includegraphics[clip, width = 0.6 \textwidth]{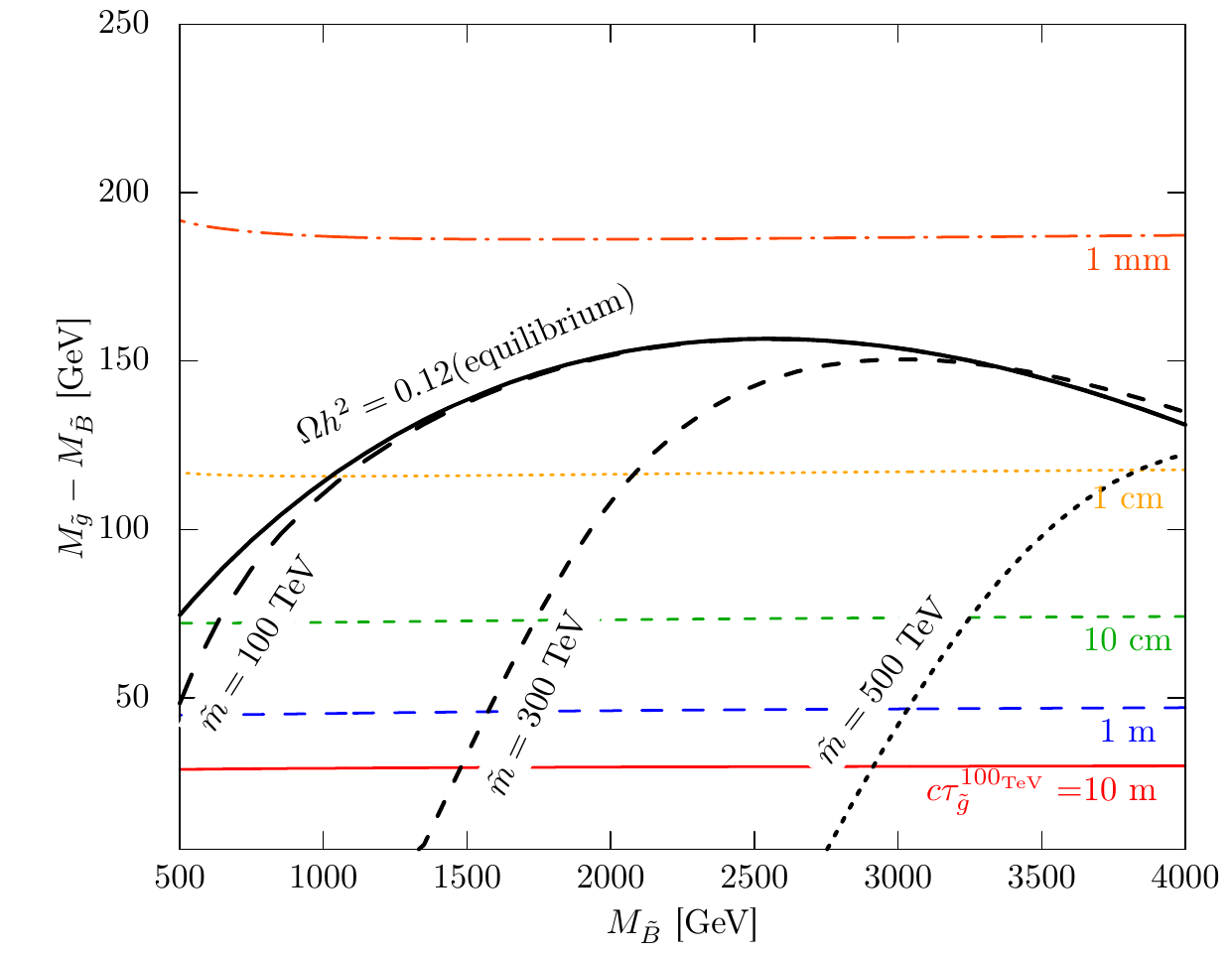}
\caption{Decay length of the gluino $c\tau^{100_{\rm TeV}}_{\tilde{g}}$
 with the squark mass $\widetilde m=100$ TeV in colored (almost
 horizontal) lines. Mass difference $\Delta M$ with which the thermal
 relic of the bino DM agrees to $\Omega_{\rm DM}h^2 = 0.12$ is also
 shown in the black solid line for the case in which the bino-gluino
 chemical equilibrium is assumed, while the cases for $\widetilde m=100$,
 300 and 500~TeV are given in the other black lines.
}
\label{fig:ctau}
\end{figure}
%%%%%%%%%%%%%%%%%%%%%%%%%%%%%%%%%%%%%%%%%%%%%%%%%%%%%%%%%%%%%%%%%

To illustrate the gluino decay length corresponding to the bino-gluino
coannihilation region, in Fig.~\ref{fig:ctau}, we plot contours of the
gluino decay length in colored lines with the squark masses set to be
$\widetilde{m} = 100$~TeV, which we denote by $c\tau^{100_{\rm
TeV}}_{\tilde{g}}$, on the $M_{\tilde{B}}$--$\Delta M$ plane. We also
show the mass difference $\Delta M$ with which the thermal relic of the
bino DM agrees to $\Omega_{\rm DM}h^2 = 0.12$; the black solid line
shows the case where the bino-gluino chemical equilibrium is assumed,
while the other black lines represent the cases of $\widetilde m=100$, 300
and 500~TeV. To avoid overproduction, $\Delta M$ should be below
these lines. From Fig.~\ref{fig:ctau}, we find that the gluino decay length is
scarcely dependent on the bino mass, which has been already shown in
Eq.~\eqref{eq:ctauglu} implicitly. We have $c \tau_{\tilde{g}}> {\cal
O}(1)$~mm where the thermal relic abundance of the bino DM explains the
observed DM density. This is a crucial observation for the strategy of
exploring the bino-gluino coannihilation region at the LHC.

%%%%%%%%%%%%%%%%%%%%%%%%%%%%%%%%%%%%%%%%%%%%%%%
\section{LHC search}
\label{sec:results}
%%%%%%%%%%%%%%%%%%%%%%%%%%%%%%%%%%%%%%%%%%%%%%%

If gluino decays promptly and the bino and gluino masses are almost
degenerate, it is quite hard to search for the gluino at the LHC, since
the small mass difference makes the missing energy and jet activities
tiny. Currently the ATLAS and CMS collaborations have put limits on
such a degenerate neutralino, \textit{i.e.}, bino in our case, with a
mass of around 600~GeV \cite{Chatrchyan:2014lfa,Aad:2014wea}. The bounds
are expected to reach $\sim 1200$~GeV with the integrated luminosity of
300 $\rm{fb}^{-1}$ at the 14 TeV LHC \cite{ATL-PHYS-PUB-2014-010}.

These limits are in fact drastically improved once we consider the fact
that in the case of the bino-gluino coannihilation
scenario, the gluino lifetime is as long as $c \tau_{\tilde{g}}> {\cal
O}(1)$~mm, as we have seen in the previous section. Such a gluino has
a distinct property in the collider experiments; a gluino with a decay
length of $c \tau_{\tilde{g}}> {\cal O}(1)$~mm leaves a visible
displaced vertex (DV) in the detectors, which greatly helps the gluino search.
At present, however, there have been no dedicated searches from this
aspect so far.\footnote{A similar discussion has been recently given in
Ref.~\cite{Liu:2015bma} based on the CMS displaced dijets results \cite{CMS:2014wda,*CMS:2013oea}, though their constraint is much weaker than
ours. As we will discuss below, the ATLAS DV search \cite{Aad:2015rba}
exploits the missing energy trigger, while the CMS search does not. In
addition, the CMS dijet search requires large scalar sum of jet
transverse momenta, which is not effective when the mass difference of
bino and gluino is small. For these reasons, at present, the ATLAS search
offers better sensitivities than the CMS one.} 

The ATLAS collaboration has searched for DVs in the
region of $|z|<30$~cm and $r<30$~cm in the inner detector
\cite{Aad:2015rba,TheATLAScollaboration:2013yia,Aad:2012zx,*Aad:2011zb}, where
$z$-axis points along the LHC beam line and $r$ 
denotes the radial coordinate in the plane perpendicular to the
$z$-axis. They use the DVs reconstructed only in the
air-gap region, namely, discard the DVs reconstructed within the material
layers. This leads to significant background reduction. The signal
region for the DVs is defined such that the number of tracks associated
with the DV is larger than four and $m_{\text{DV}}>10$~GeV, where
$m_{\text{DV}}$ is the invariant mass of the tracks evaluated with the
charged-pion mass hypothesis. Since they have observed no event in the
signal region, they have given an upper limit on the long-lived gluino
production cross section, which is interpreted as bound on the gluino mass
in the high-scale SUSY scenario with a fixed neutralino mass of
100~GeV \cite{Aad:2015rba}.

%%%%%%%%%%%%%%%%%%%%%%%%%%%%%%%%%%%%%%%%%%%%%%%%%%%%%%%%%%%%%%%%%%%%%%%%%%
\begin{figure}[t]
\centering
\includegraphics[clip, width = 0.6 \textwidth]{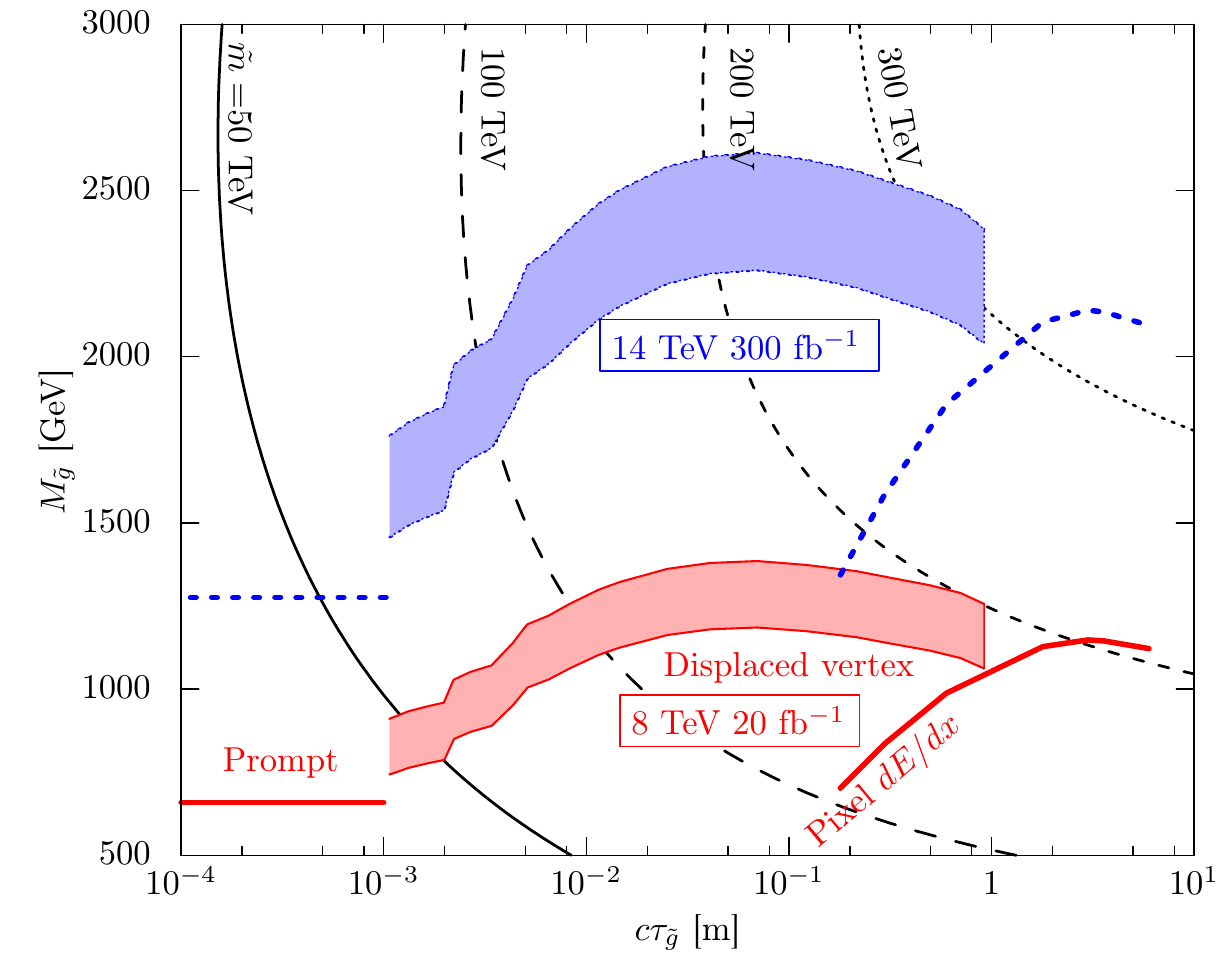}
\caption{Current constraints (red and solid lines) and future prospects (blue and dashed lines) for the gluino searches. Favored region for the DM
 relic abundance is also shown in black lines for the cases of
 $\widetilde{m}=50$, 100, 200, and 300~TeV, with $\Delta M$ chosen so that
 the thermal relic abundance of the bino equals to the current observed
 DM density. We also show the current constraint \cite{Aad:2014wea} and
 future prospect of the 14~TeV LHC run \cite{ATL-PHYS-PUB-2014-010} from
 the search for the prompt-decay gluino in horizontal red solid and blue
 dashed lines, respectively. }
\label{fig:prospect}
\end{figure}
%%%%%%%%%%%%%%%%%%%%%%%%%%%%%%%%%%%%%%%%%%%%%%%%%%%%%%%%%%%%%%%%%%%%%%%%

We re-interpret this low $m_{\text{DV}}$ search result in the case
of the degenerate bino-gluino system, and obtain constraints on the
bino-gluino coannihilation scenario, which is shown in
Fig.~\ref{fig:prospect}. Here, the red and blue bands (from
$c\tau_{\tilde{g}} = 1$~mm to 1~m) show the estimated sensitivities of
the DV search with the total luminosity of 20~$\rm{fb}^{-1}$ at the
8~TeV running and with 300~$\rm{fb}^{-1}$ at 14~TeV, respectively. The
upper lines of these bands are for the cases where only the trigger
efficiency is taken into account, which are simulated with
\textsc{HERWIG 6} \cite{Corcella:2000bw,*Corcella:2002jc} and
\textsc{AcerDET} \cite{RichterWas:2002ch} to be 40\% for
8~TeV with the threshold of the missing energy of 100~GeV, and 15\% for
14~TeV with the missing energy trigger of 200~GeV. Their dependence on
the mass of gluino is only a few percent level. The lower lines, on the
other hand, correspond to the reconstruction efficiency for DVs that
is estimated from Refs.~\cite{TheATLAScollaboration:2013yia}, where the
long-lived neutralino decaying to two quarks and 
one muon is discussed in the $R$-parity violating SUSY scenario.
The reconstruction efficiency for the 108~GeV neutralino is about 20\%
of that for the 494~GeV neutralino in this case; we use this 20\% for the lower
lines, which gives conservative limits rather than the previous ones. 
In Fig.~\ref{fig:prospect}, we also show the favored region in terms of
the DM relic abundance in black lines for the cases of
$\widetilde{m}=50$, 100, 200, and 300~TeV. Here, the bino-gluino mass
difference $\Delta M$ is taken such that the thermal relic of bino DM
explains the correct DM density. This reads that the present LHC data
have already constrained a considerable range of parameter region
consistent with the bino-gluino coannihilation scenario. This constraint
is in fact much stronger than the ordinary limit from the searches of
promptly decaying gluinos, which are based on only jets and missing
energy \cite{Chatrchyan:2014lfa,Aad:2014wea}. 
This constraint is indicated by the red and solid horizontal line in
this figure. The 14~TeV LHC 
running can further probe this scenario and reach $M_{\tilde{g}}\sim
2.5$~TeV when $c\tau_{\tilde{g}}={\cal O}(1-10)$~cm; this sensitivity is
better than that by search with only jets and missing energy
\cite{ATL-PHYS-PUB-2014-010} (shown in the horizontal blue dashed line
in the above figure) by almost a factor of two. 

In addition, the ATLAS collaboration searches for massive charged meta-stable
particles, such as $R$-hadrons \cite{Farrar:1978xj}, with an another
approach \cite{ATLAS-CONF-2015-013}.  A characteristic
feature of such particles is that they are 
produced with relatively low velocities, $\beta \equiv v/c <1$. This
signature can be seen by means of large energy loss, $dE/dx$, in the
ATLAS Pixel detector. Here, we note that this analysis requires gluinos
to form charged $R$-hadrons. Although the estimation of the charged
hadronization fraction of gluinos may suffer from large theoretical
uncertainty, this search offers the best sensitivity for
$c\tau_{\tilde{g}}> 1$~m. In Ref.~\cite{ATLAS-CONF-2015-013}, the result
of this search is given as limits on the gluino mass in the case of
$\Delta M = 100$~GeV. We use the trigger efficiency given there for our
computation for the 8~TeV case, and estimate the efficiency for the
14~TeV case by re-scaling it with a factor obtained by simulations.
The red and blue solid curves in Fig.~\ref{fig:prospect} show the
estimated sensitivities of this search with 20~$\rm{fb}^{-1}$ at 8~TeV
and with 300~$\rm{fb}^{-1}$ at 14~TeV, respectively. We find that
the searches of heavy stable charged particles give the most stringent
constraints when $c\tau_{\tilde{g}}>1$~m, and are complementary to the
DV searches. In particular, they are of importance when the scalar mass
scale is relatively higher, say, a few hundred~TeV.

%%%%%%%%%%%%%%%%%%%%%%%%%%%%%%%%%%%%%%%%
\section{Conclusion and discussion}
\label{sec:conclusion}
%%%%%%%%%%%%%%%%%%%%%%%%%%%%%%%%%%%%%%%%

In this paper, we study the bino-gluino coannihilation in the high-scale
SUSY scenario. We have found that the squark mass scale cannot be too
large for the coannihilation to work well. The upper bound on the squark
mass is 200--1000~TeV for the gluino mass 1--8~TeV. Actually this mass
scale is coincident with the prediction of the spectrum often called
the spread or mini-split SUSY \cite{Wells:2003tf,*Wells:2004di,
ArkaniHamed:2004fb,*Giudice:2004tc,*ArkaniHamed:2004yi,*ArkaniHamed:2005yv,
Hall:2011jd, Hall:2012zp, Ibe:2006de, *Ibe:2011aa, *Ibe:2012hu,
Arvanitaki:2012ps, ArkaniHamed:2012gw, Evans:2013lpa, *Evans:2013dza,
*Evans:2013uza,*Evans:2014hda, Nomura:2014asa}.
This constraint will provide a new perspective on the model-building to
realize such mass spectrum.

We also discuss the LHC signatures of this scenario. Because of the small mass
difference between the bino LSP and gluino, which is necessary for 
coannihilation, as well as heavy squark masses, the gluino decay length is
considerably prolonged. Despite the small jets and missing energy
activity, the DV and $R$-hadron searches can efficiently probe such
long-lived gluinos. If the squark mass scale is higher than about
100~TeV, the current lower bound on the gluino mass is around
1.2~TeV. The 13/14 TeV LHC 300~fb$^{-1}$ stage is expected to be able to explore
gluinos with a mass of $\sim 2$~TeV.

Let us speculate possible sensitivities for much higher energy machines.
For gluinos with the decay length longer than ${\cal O}(1)$~mm, a mass
of 4.5 (10)~TeV can be probed using a $\sqrt{s}=33~(100)$~TeV running
proton collider with the integrated luminosity of 100 fb$^{-1}$, provided
that the background is sufficiently small and the detection efficiency of
gluinos is the same as that of the current LHC detector. 
This estimation may, of course, be too naive. Further detailed studies
should be dedicated to see more precise prospects for such
colliders, though we expect that they can probe the most of parameter
space of the bino-gluino coannihilation.

Lastly, we discuss the possibility of other gaugino coannihilation scenarios.
As in the case of the current study, the small mass difference and
heavier sfermion scale easily make the next LSP live long. For instance,
in the case of the wino and gluino coannihilation, we may observe very
exotic signatures; if the gluino lifetime is long enough, the gluino can
carry the charged wino to the LHC trackers. In this case, we may observe
displaced and disappearing tracks of the charged wino. The
large gluino production cross section and the long-lived
nature of the charged wino make it rather easy to look for this scenario
in the LHC experiments. Another very interesting and plausible
possibility is bino-wino coannihilation. This spectrum can be relatively
easily realized even in the minimal anomaly mediation model. In this
case, we may have another long-lived particle, which may play an important
role at the LHC searches. A detailed analysis for this scenario will be
done elsewhere \cite{WIP}.

%%%%%%%%%%%%%%%%%%%%%%%%%%%%%%%%%%%%
\section*{Acknowledgments}
%%%%%%%%%%%%%%%%%%%%%%%%%%%%%%%%%%%%

N.N. thanks Jason L. Evans and Keith A. Olive for useful discussions. 
The work of N.N. is supported by Research Fellowships of the Japan Society
for the Promotion of Science for Young Scientists.

%%%%%%%%%%%%%%%%%%%%%%%%%%%%%%%%%%%%%%%%%%%%%%
%\section*{Appendix}
%\appendix
%%%%%%%%%%%%%%%%%%%%%%%%%%%%%%%%%%%%%%%%%%%%%

%%%%%%%%%%%%% References %%%%%%%%%%%%%%%%%%%
\bibliographystyle{aps}
\bibliography{ref}

%merlin.mbs apsrev4-1.bst 2010-07-25 4.21a (PWD, AO, DPC) hacked
%Control: key (0)
%Control: author (72) initials jnrlst
%Control: editor formatted (1) identically to author
%Control: production of article title (-1) disabled
%Control: page (0) single
%Control: year (1) truncated
%Control: production of eprint (0) enabled
\begin{thebibliography}{106}%
\makeatletter
\providecommand \@ifxundefined [1]{%
 \@ifx{#1\undefined}
}%
\providecommand \@ifnum [1]{%
 \ifnum #1\expandafter \@firstoftwo
 \else \expandafter \@secondoftwo
 \fi
}%
\providecommand \@ifx [1]{%
 \ifx #1\expandafter \@firstoftwo
 \else \expandafter \@secondoftwo
 \fi
}%
\providecommand \natexlab [1]{#1}%
\providecommand \enquote  [1]{``#1''}%
\providecommand \bibnamefont  [1]{#1}%
\providecommand \bibfnamefont [1]{#1}%
\providecommand \citenamefont [1]{#1}%
\providecommand \href@noop [0]{\@secondoftwo}%
\providecommand \href [0]{\begingroup \@sanitize@url \@href}%
\providecommand \@href[1]{\@@startlink{#1}\@@href}%
\providecommand \@@href[1]{\endgroup#1\@@endlink}%
\providecommand \@sanitize@url [0]{\catcode `\\12\catcode `\$12\catcode
  `\&12\catcode `\#12\catcode `\^12\catcode `\_12\catcode `\%12\relax}%
\providecommand \@@startlink[1]{}%
\providecommand \@@endlink[0]{}%
\providecommand \url  [0]{\begingroup\@sanitize@url \@url }%
\providecommand \@url [1]{\endgroup\@href {#1}{\urlprefix }}%
\providecommand \urlprefix  [0]{URL }%
\providecommand \Eprint [0]{\href }%
\providecommand \doibase [0]{http://dx.doi.org/}%
\providecommand \selectlanguage [0]{\@gobble}%
\providecommand \bibinfo  [0]{\@secondoftwo}%
\providecommand \bibfield  [0]{\@secondoftwo}%
\providecommand \translation [1]{[#1]}%
\providecommand \BibitemOpen [0]{}%
\providecommand \bibitemStop [0]{}%
\providecommand \bibitemNoStop [0]{.\EOS\space}%
\providecommand \EOS [0]{\spacefactor3000\relax}%
\providecommand \BibitemShut  [1]{\csname bibitem#1\endcsname}%
\let\auto@bib@innerbib\@empty
%</preamble>
\bibitem [{\citenamefont {Aad}\ \emph {et~al.}(2012{\natexlab{a}})\citenamefont
  {Aad} \emph {et~al.}}]{Aad:2012tfa}%
  \BibitemOpen
  \bibfield  {author} {\bibinfo {author} {\bibfnamefont {G.}~\bibnamefont
  {Aad}} \emph {et~al.} (\bibinfo {collaboration} {ATLAS}),\ }\href {\doibase
  10.1016/j.physletb.2012.08.020} {\bibfield  {journal} {\bibinfo  {journal}
  {Phys.Lett.}\ }\textbf {\bibinfo {volume} {B716}},\ \bibinfo {pages} {1}
  (\bibinfo {year} {2012}{\natexlab{a}})},\ \Eprint
  {http://arxiv.org/abs/1207.7214}{arXiv:1207.7214 [hep-ex]}\BibitemShut
  {NoStop}%
%%CITATION = ARXIV:1207.7214;%%
\bibitem [{\citenamefont {Chatrchyan}\ \emph {et~al.}(2012)\citenamefont
  {Chatrchyan} \emph {et~al.}}]{Chatrchyan:2012ufa}%
  \BibitemOpen
  \bibfield  {author} {\bibinfo {author} {\bibfnamefont {S.}~\bibnamefont
  {Chatrchyan}} \emph {et~al.} (\bibinfo {collaboration} {CMS}),\ }\href
  {\doibase 10.1016/j.physletb.2012.08.021} {\bibfield  {journal} {\bibinfo
  {journal} {Phys.Lett.}\ }\textbf {\bibinfo {volume} {B716}},\ \bibinfo
  {pages} {30} (\bibinfo {year} {2012})},\ \Eprint
  {http://arxiv.org/abs/1207.7235}{arXiv:1207.7235 [hep-ex]}\BibitemShut
  {NoStop}%
%%CITATION = ARXIV:1207.7235;%%
\bibitem [{\citenamefont {Aad}\ \emph {et~al.}(2015{\natexlab{a}})\citenamefont
  {Aad} \emph {et~al.}}]{Aad:2015zhl}%
  \BibitemOpen
  \bibfield  {author} {\bibinfo {author} {\bibfnamefont {G.}~\bibnamefont
  {Aad}} \emph {et~al.} (\bibinfo {collaboration} {ATLAS, CMS}),\ }\href
  {\doibase 10.1103/PhysRevLett.114.191803} {\bibfield  {journal} {\bibinfo
  {journal} {Phys.Rev.Lett.}\ }\textbf {\bibinfo {volume} {114}},\ \bibinfo
  {pages} {191803} (\bibinfo {year} {2015}{\natexlab{a}})},\ \Eprint
  {http://arxiv.org/abs/1503.07589}{arXiv:1503.07589 [hep-ex]}\BibitemShut
  {NoStop}%
%%CITATION = ARXIV:1503.07589;%%
\bibitem [{\citenamefont {Inoue}\ \emph {et~al.}(1982)\citenamefont {Inoue},
  \citenamefont {Kakuto}, \citenamefont {Komatsu},\ and\ \citenamefont
  {Takeshita}}]{Inoue:1982ej}%
  \BibitemOpen
  \bibfield  {author} {\bibinfo {author} {\bibfnamefont {K.}~\bibnamefont
  {Inoue}}, \bibinfo {author} {\bibfnamefont {A.}~\bibnamefont {Kakuto}},
  \bibinfo {author} {\bibfnamefont {H.}~\bibnamefont {Komatsu}}, \ and\
  \bibinfo {author} {\bibfnamefont {S.}~\bibnamefont {Takeshita}},\ }\href
  {\doibase 10.1143/PTP.67.1889} {\bibfield  {journal} {\bibinfo  {journal}
  {Prog.Theor.Phys.}\ }\textbf {\bibinfo {volume} {67}},\ \bibinfo {pages}
  {1889} (\bibinfo {year} {1982})}\BibitemShut {NoStop}%
%%CITATION = PTPKA,67,1889;%%
\bibitem [{\citenamefont {Flores}\ and\ \citenamefont
  {Sher}(1983)}]{Flores:1982pr}%
  \BibitemOpen
  \bibfield  {author} {\bibinfo {author} {\bibfnamefont {R.~A.}\ \bibnamefont
  {Flores}}\ and\ \bibinfo {author} {\bibfnamefont {M.}~\bibnamefont {Sher}},\
  }\href {\doibase 10.1016/0003-4916(83)90331-7} {\bibfield  {journal}
  {\bibinfo  {journal} {Annals Phys.}\ }\textbf {\bibinfo {volume} {148}},\
  \bibinfo {pages} {95} (\bibinfo {year} {1983})}\BibitemShut {NoStop}%
%%CITATION = APNYA,148,95;%%
\bibitem [{\citenamefont {Okada}\ \emph
  {et~al.}(1991{\natexlab{a}})\citenamefont {Okada}, \citenamefont
  {Yamaguchi},\ and\ \citenamefont {Yanagida}}]{Okada:1990vk}%
  \BibitemOpen
  \bibfield  {author} {\bibinfo {author} {\bibfnamefont {Y.}~\bibnamefont
  {Okada}}, \bibinfo {author} {\bibfnamefont {M.}~\bibnamefont {Yamaguchi}}, \
  and\ \bibinfo {author} {\bibfnamefont {T.}~\bibnamefont {Yanagida}},\ }\href
  {\doibase 10.1143/PTP.85.1} {\bibfield  {journal} {\bibinfo  {journal}
  {Prog.Theor.Phys.}\ }\textbf {\bibinfo {volume} {85}},\ \bibinfo {pages} {1}
  (\bibinfo {year} {1991}{\natexlab{a}})}\BibitemShut {NoStop}%
%%CITATION = PTPKA,85,1;%%
\bibitem [{\citenamefont {Okada}\ \emph
  {et~al.}(1991{\natexlab{b}})\citenamefont {Okada}, \citenamefont
  {Yamaguchi},\ and\ \citenamefont {Yanagida}}]{Okada:1990gg}%
  \BibitemOpen
  \bibfield  {author} {\bibinfo {author} {\bibfnamefont {Y.}~\bibnamefont
  {Okada}}, \bibinfo {author} {\bibfnamefont {M.}~\bibnamefont {Yamaguchi}}, \
  and\ \bibinfo {author} {\bibfnamefont {T.}~\bibnamefont {Yanagida}},\ }\href
  {\doibase 10.1016/0370-2693(91)90642-4} {\bibfield  {journal} {\bibinfo
  {journal} {Phys.Lett.}\ }\textbf {\bibinfo {volume} {B262}},\ \bibinfo
  {pages} {54} (\bibinfo {year} {1991}{\natexlab{b}})}\BibitemShut {NoStop}%
%%CITATION = PHLTA,B262,54;%%
\bibitem [{\citenamefont {Ellis}\ \emph
  {et~al.}(1991{\natexlab{a}})\citenamefont {Ellis}, \citenamefont {Ridolfi},\
  and\ \citenamefont {Zwirner}}]{Ellis:1990nz}%
  \BibitemOpen
  \bibfield  {author} {\bibinfo {author} {\bibfnamefont {J.~R.}\ \bibnamefont
  {Ellis}}, \bibinfo {author} {\bibfnamefont {G.}~\bibnamefont {Ridolfi}}, \
  and\ \bibinfo {author} {\bibfnamefont {F.}~\bibnamefont {Zwirner}},\ }\href
  {\doibase 10.1016/0370-2693(91)90863-L} {\bibfield  {journal} {\bibinfo
  {journal} {Phys.Lett.}\ }\textbf {\bibinfo {volume} {B257}},\ \bibinfo
  {pages} {83} (\bibinfo {year} {1991}{\natexlab{a}})}\BibitemShut {NoStop}%
%%CITATION = PHLTA,B257,83;%%
\bibitem [{\citenamefont {Haber}\ and\ \citenamefont
  {Hempfling}(1991)}]{Haber:1990aw}%
  \BibitemOpen
  \bibfield  {author} {\bibinfo {author} {\bibfnamefont {H.~E.}\ \bibnamefont
  {Haber}}\ and\ \bibinfo {author} {\bibfnamefont {R.}~\bibnamefont
  {Hempfling}},\ }\href {\doibase 10.1103/PhysRevLett.66.1815} {\bibfield
  {journal} {\bibinfo  {journal} {Phys.Rev.Lett.}\ }\textbf {\bibinfo {volume}
  {66}},\ \bibinfo {pages} {1815} (\bibinfo {year} {1991})}\BibitemShut
  {NoStop}%
%%CITATION = PRLTA,66,1815;%%
\bibitem [{\citenamefont {Ellis}\ \emph
  {et~al.}(1991{\natexlab{b}})\citenamefont {Ellis}, \citenamefont {Ridolfi},\
  and\ \citenamefont {Zwirner}}]{Ellis:1991zd}%
  \BibitemOpen
  \bibfield  {author} {\bibinfo {author} {\bibfnamefont {J.~R.}\ \bibnamefont
  {Ellis}}, \bibinfo {author} {\bibfnamefont {G.}~\bibnamefont {Ridolfi}}, \
  and\ \bibinfo {author} {\bibfnamefont {F.}~\bibnamefont {Zwirner}},\ }\href
  {\doibase 10.1016/0370-2693(91)90626-2} {\bibfield  {journal} {\bibinfo
  {journal} {Phys.Lett.}\ }\textbf {\bibinfo {volume} {B262}},\ \bibinfo
  {pages} {477} (\bibinfo {year} {1991}{\natexlab{b}})}\BibitemShut {NoStop}%
%%CITATION = PHLTA,B262,477;%%
\bibitem [{\citenamefont {Giudice}\ and\ \citenamefont
  {Strumia}(2012)}]{Giudice:2011cg}%
  \BibitemOpen
  \bibfield  {author} {\bibinfo {author} {\bibfnamefont {G.~F.}\ \bibnamefont
  {Giudice}}\ and\ \bibinfo {author} {\bibfnamefont {A.}~\bibnamefont
  {Strumia}},\ }\href {\doibase 10.1016/j.nuclphysb.2012.01.001} {\bibfield
  {journal} {\bibinfo  {journal} {Nucl.Phys.}\ }\textbf {\bibinfo {volume}
  {B858}},\ \bibinfo {pages} {63} (\bibinfo {year} {2012})},\ \Eprint
  {http://arxiv.org/abs/1108.6077}{arXiv:1108.6077 [hep-ph]}\BibitemShut
  {NoStop}%
%%CITATION = ARXIV:1108.6077;%%
\bibitem [{\citenamefont {Aad}\ \emph {et~al.}(2014)\citenamefont {Aad} \emph
  {et~al.}}]{Aad:2014wea}%
  \BibitemOpen
  \bibfield  {author} {\bibinfo {author} {\bibfnamefont {G.}~\bibnamefont
  {Aad}} \emph {et~al.} (\bibinfo {collaboration} {ATLAS}),\ }\href {\doibase
  10.1007/JHEP09(2014)176} {\bibfield  {journal} {\bibinfo  {journal} {JHEP}\
  }\textbf {\bibinfo {volume} {1409}},\ \bibinfo {pages} {176} (\bibinfo {year}
  {2014})},\ \Eprint {http://arxiv.org/abs/1405.7875}{arXiv:1405.7875
  [hep-ex]}\BibitemShut {NoStop}%
%%CITATION = ARXIV:1405.7875;%%
\bibitem [{\citenamefont {Chatrchyan}\ \emph {et~al.}(2014)\citenamefont
  {Chatrchyan} \emph {et~al.}}]{Chatrchyan:2014lfa}%
  \BibitemOpen
  \bibfield  {author} {\bibinfo {author} {\bibfnamefont {S.}~\bibnamefont
  {Chatrchyan}} \emph {et~al.} (\bibinfo {collaboration} {CMS}),\ }\href
  {\doibase 10.1007/JHEP06(2014)055} {\bibfield  {journal} {\bibinfo  {journal}
  {JHEP}\ }\textbf {\bibinfo {volume} {1406}},\ \bibinfo {pages} {055}
  (\bibinfo {year} {2014})},\ \Eprint
  {http://arxiv.org/abs/1402.4770}{arXiv:1402.4770 [hep-ex]}\BibitemShut
  {NoStop}%
%%CITATION = ARXIV:1402.4770;%%
\bibitem [{\citenamefont {Gabbiani}\ \emph {et~al.}(1996)\citenamefont
  {Gabbiani}, \citenamefont {Gabrielli}, \citenamefont {Masiero},\ and\
  \citenamefont {Silvestrini}}]{Gabbiani:1996hi}%
  \BibitemOpen
  \bibfield  {author} {\bibinfo {author} {\bibfnamefont {F.}~\bibnamefont
  {Gabbiani}}, \bibinfo {author} {\bibfnamefont {E.}~\bibnamefont {Gabrielli}},
  \bibinfo {author} {\bibfnamefont {A.}~\bibnamefont {Masiero}}, \ and\
  \bibinfo {author} {\bibfnamefont {L.}~\bibnamefont {Silvestrini}},\ }\href
  {\doibase 10.1016/0550-3213(96)00390-2} {\bibfield  {journal} {\bibinfo
  {journal} {Nucl.Phys.}\ }\textbf {\bibinfo {volume} {B477}},\ \bibinfo
  {pages} {321} (\bibinfo {year} {1996})},\ \Eprint
  {http://arxiv.org/abs/hep-ph/9604387}{arXiv:hep-ph/9604387
  [hep-ph]}\BibitemShut {NoStop}%
%%CITATION = HEP-PH/9604387;%%
\bibitem [{\citenamefont {Moroi}\ and\ \citenamefont
  {Nagai}(2013)}]{Moroi:2013sfa}%
  \BibitemOpen
  \bibfield  {author} {\bibinfo {author} {\bibfnamefont {T.}~\bibnamefont
  {Moroi}}\ and\ \bibinfo {author} {\bibfnamefont {M.}~\bibnamefont {Nagai}},\
  }\href {\doibase 10.1016/j.physletb.2013.04.049} {\bibfield  {journal}
  {\bibinfo  {journal} {Phys.Lett.}\ }\textbf {\bibinfo {volume} {B723}},\
  \bibinfo {pages} {107} (\bibinfo {year} {2013})},\ \Eprint
  {http://arxiv.org/abs/1303.0668}{arXiv:1303.0668 [hep-ph]}\BibitemShut
  {NoStop}%
%%CITATION = ARXIV:1303.0668;%%
\bibitem [{\citenamefont {McKeen}\ \emph {et~al.}(2013)\citenamefont {McKeen},
  \citenamefont {Pospelov},\ and\ \citenamefont {Ritz}}]{McKeen:2013dma}%
  \BibitemOpen
  \bibfield  {author} {\bibinfo {author} {\bibfnamefont {D.}~\bibnamefont
  {McKeen}}, \bibinfo {author} {\bibfnamefont {M.}~\bibnamefont {Pospelov}}, \
  and\ \bibinfo {author} {\bibfnamefont {A.}~\bibnamefont {Ritz}},\ }\href
  {\doibase 10.1103/PhysRevD.87.113002} {\bibfield  {journal} {\bibinfo
  {journal} {Phys.Rev.}\ }\textbf {\bibinfo {volume} {D87}},\ \bibinfo {pages}
  {113002} (\bibinfo {year} {2013})},\ \Eprint
  {http://arxiv.org/abs/1303.1172}{arXiv:1303.1172 [hep-ph]}\BibitemShut
  {NoStop}%
%%CITATION = ARXIV:1303.1172;%%
\bibitem [{\citenamefont {Altmannshofer}\ \emph {et~al.}(2013)\citenamefont
  {Altmannshofer}, \citenamefont {Harnik},\ and\ \citenamefont
  {Zupan}}]{Altmannshofer:2013lfa}%
  \BibitemOpen
  \bibfield  {author} {\bibinfo {author} {\bibfnamefont {W.}~\bibnamefont
  {Altmannshofer}}, \bibinfo {author} {\bibfnamefont {R.}~\bibnamefont
  {Harnik}}, \ and\ \bibinfo {author} {\bibfnamefont {J.}~\bibnamefont
  {Zupan}},\ }\href {\doibase 10.1007/JHEP11(2013)202} {\bibfield  {journal}
  {\bibinfo  {journal} {JHEP}\ }\textbf {\bibinfo {volume} {1311}},\ \bibinfo
  {pages} {202} (\bibinfo {year} {2013})},\ \Eprint
  {http://arxiv.org/abs/1308.3653}{arXiv:1308.3653 [hep-ph]}\BibitemShut
  {NoStop}%
%%CITATION = ARXIV:1308.3653;%%
\bibitem [{\citenamefont {Fuyuto}\ \emph {et~al.}(2013)\citenamefont {Fuyuto},
  \citenamefont {Hisano}, \citenamefont {Nagata},\ and\ \citenamefont
  {Tsumura}}]{Fuyuto:2013gla}%
  \BibitemOpen
  \bibfield  {author} {\bibinfo {author} {\bibfnamefont {K.}~\bibnamefont
  {Fuyuto}}, \bibinfo {author} {\bibfnamefont {J.}~\bibnamefont {Hisano}},
  \bibinfo {author} {\bibfnamefont {N.}~\bibnamefont {Nagata}}, \ and\ \bibinfo
  {author} {\bibfnamefont {K.}~\bibnamefont {Tsumura}},\ }\href {\doibase
  10.1007/JHEP12(2013)010} {\bibfield  {journal} {\bibinfo  {journal} {JHEP}\
  }\textbf {\bibinfo {volume} {1312}},\ \bibinfo {pages} {010} (\bibinfo {year}
  {2013})},\ \Eprint {http://arxiv.org/abs/1308.6493}{arXiv:1308.6493
  [hep-ph]}\BibitemShut {NoStop}%
%%CITATION = ARXIV:1308.6493;%%
\bibitem [{\citenamefont {Tanimoto}\ and\ \citenamefont
  {Yamamoto}(2014)}]{Tanimoto:2014eva}%
  \BibitemOpen
  \bibfield  {author} {\bibinfo {author} {\bibfnamefont {M.}~\bibnamefont
  {Tanimoto}}\ and\ \bibinfo {author} {\bibfnamefont {K.}~\bibnamefont
  {Yamamoto}},\ }\href {\doibase 10.1016/j.physletb.2014.06.067} {\bibfield
  {journal} {\bibinfo  {journal} {Phys.Lett.}\ }\textbf {\bibinfo {volume}
  {B735}},\ \bibinfo {pages} {426} (\bibinfo {year} {2014})},\ \Eprint
  {http://arxiv.org/abs/1404.0520}{arXiv:1404.0520 [hep-ph]}\BibitemShut
  {NoStop}%
%%CITATION = ARXIV:1404.0520;%%
\bibitem [{\citenamefont {Tanimoto}\ and\ \citenamefont
  {Yamamoto}(2015)}]{Tanimoto:2015ota}%
  \BibitemOpen
  \bibfield  {author} {\bibinfo {author} {\bibfnamefont {M.}~\bibnamefont
  {Tanimoto}}\ and\ \bibinfo {author} {\bibfnamefont {K.}~\bibnamefont
  {Yamamoto}},\ }\href {\doibase 10.1093/ptep/ptv066} {\bibfield  {journal}
  {\bibinfo  {journal} {Prog.Theor.Exp.Phys.}\ ,\ \bibinfo {pages} {053B07}}
  (\bibinfo {year} {2015})},\ \Eprint
  {http://arxiv.org/abs/1503.06270}{arXiv:1503.06270 [hep-ph]}\BibitemShut
  {NoStop}%
%%CITATION = ARXIV:1503.06270;%%
\bibitem [{\citenamefont {Liu}\ and\ \citenamefont {Nath}(2013)}]{Liu:2013ula}%
  \BibitemOpen
  \bibfield  {author} {\bibinfo {author} {\bibfnamefont {M.}~\bibnamefont
  {Liu}}\ and\ \bibinfo {author} {\bibfnamefont {P.}~\bibnamefont {Nath}},\
  }\href {\doibase 10.1103/PhysRevD.87.095012} {\bibfield  {journal} {\bibinfo
  {journal} {Phys.Rev.}\ }\textbf {\bibinfo {volume} {D87}},\ \bibinfo {pages}
  {095012} (\bibinfo {year} {2013})},\ \Eprint
  {http://arxiv.org/abs/1303.7472}{arXiv:1303.7472 [hep-ph]}\BibitemShut
  {NoStop}%
%%CITATION = ARXIV:1303.7472;%%
\bibitem [{\citenamefont {Hisano}\ \emph
  {et~al.}(2013{\natexlab{a}})\citenamefont {Hisano}, \citenamefont
  {Kobayashi}, \citenamefont {Kuwahara},\ and\ \citenamefont
  {Nagata}}]{Hisano:2013exa}%
  \BibitemOpen
  \bibfield  {author} {\bibinfo {author} {\bibfnamefont {J.}~\bibnamefont
  {Hisano}}, \bibinfo {author} {\bibfnamefont {D.}~\bibnamefont {Kobayashi}},
  \bibinfo {author} {\bibfnamefont {T.}~\bibnamefont {Kuwahara}}, \ and\
  \bibinfo {author} {\bibfnamefont {N.}~\bibnamefont {Nagata}},\ }\href
  {\doibase 10.1007/JHEP07(2013)038} {\bibfield  {journal} {\bibinfo  {journal}
  {JHEP}\ }\textbf {\bibinfo {volume} {1307}},\ \bibinfo {pages} {038}
  (\bibinfo {year} {2013}{\natexlab{a}})},\ \Eprint
  {http://arxiv.org/abs/1304.3651}{arXiv:1304.3651 [hep-ph]}\BibitemShut
  {NoStop}%
%%CITATION = ARXIV:1304.3651;%%
\bibitem [{\citenamefont {Dine}\ \emph {et~al.}(2014)\citenamefont {Dine},
  \citenamefont {Draper},\ and\ \citenamefont {Shepherd}}]{Dine:2013nga}%
  \BibitemOpen
  \bibfield  {author} {\bibinfo {author} {\bibfnamefont {M.}~\bibnamefont
  {Dine}}, \bibinfo {author} {\bibfnamefont {P.}~\bibnamefont {Draper}}, \ and\
  \bibinfo {author} {\bibfnamefont {W.}~\bibnamefont {Shepherd}},\ }\href
  {\doibase 10.1007/JHEP02(2014)027} {\bibfield  {journal} {\bibinfo  {journal}
  {JHEP}\ }\textbf {\bibinfo {volume} {1402}},\ \bibinfo {pages} {027}
  (\bibinfo {year} {2014})},\ \Eprint
  {http://arxiv.org/abs/1308.0274}{arXiv:1308.0274 [hep-ph]}\BibitemShut
  {NoStop}%
%%CITATION = ARXIV:1308.0274;%%
\bibitem [{\citenamefont {Nagata}\ and\ \citenamefont
  {Shirai}(2014)}]{Nagata:2013sba}%
  \BibitemOpen
  \bibfield  {author} {\bibinfo {author} {\bibfnamefont {N.}~\bibnamefont
  {Nagata}}\ and\ \bibinfo {author} {\bibfnamefont {S.}~\bibnamefont
  {Shirai}},\ }\href {\doibase 10.1007/JHEP03(2014)049} {\bibfield  {journal}
  {\bibinfo  {journal} {JHEP}\ }\textbf {\bibinfo {volume} {1403}},\ \bibinfo
  {pages} {049} (\bibinfo {year} {2014})},\ \Eprint
  {http://arxiv.org/abs/1312.7854}{arXiv:1312.7854 [hep-ph]}\BibitemShut
  {NoStop}%
%%CITATION = ARXIV:1312.7854;%%
\bibitem [{\citenamefont {Hall}\ \emph {et~al.}(2014)\citenamefont {Hall},
  \citenamefont {Nomura},\ and\ \citenamefont {Shirai}}]{Hall:2014vga}%
  \BibitemOpen
  \bibfield  {author} {\bibinfo {author} {\bibfnamefont {L.~J.}\ \bibnamefont
  {Hall}}, \bibinfo {author} {\bibfnamefont {Y.}~\bibnamefont {Nomura}}, \ and\
  \bibinfo {author} {\bibfnamefont {S.}~\bibnamefont {Shirai}},\ }\href
  {\doibase 10.1007/JHEP06(2014)137} {\bibfield  {journal} {\bibinfo  {journal}
  {JHEP}\ }\textbf {\bibinfo {volume} {1406}},\ \bibinfo {pages} {137}
  (\bibinfo {year} {2014})},\ \Eprint
  {http://arxiv.org/abs/1403.8138}{arXiv:1403.8138 [hep-ph]}\BibitemShut
  {NoStop}%
%%CITATION = ARXIV:1403.8138;%%
\bibitem [{\citenamefont {Evans}\ \emph
  {et~al.}(2015{\natexlab{a}})\citenamefont {Evans}, \citenamefont {Nagata},\
  and\ \citenamefont {Olive}}]{Evans:2015bxa}%
  \BibitemOpen
  \bibfield  {author} {\bibinfo {author} {\bibfnamefont {J.~L.}\ \bibnamefont
  {Evans}}, \bibinfo {author} {\bibfnamefont {N.}~\bibnamefont {Nagata}}, \
  and\ \bibinfo {author} {\bibfnamefont {K.~A.}\ \bibnamefont {Olive}},\ }\href
  {\doibase 10.1103/PhysRevD.91.055027} {\bibfield  {journal} {\bibinfo
  {journal} {Phys.Rev.}\ }\textbf {\bibinfo {volume} {D91}},\ \bibinfo {pages}
  {055027} (\bibinfo {year} {2015}{\natexlab{a}})},\ \Eprint
  {http://arxiv.org/abs/1502.00034}{arXiv:1502.00034 [hep-ph]}\BibitemShut
  {NoStop}%
%%CITATION = ARXIV:1502.00034;%%
\bibitem [{\citenamefont {Sakai}(1981)}]{Sakai:1981gr}%
  \BibitemOpen
  \bibfield  {author} {\bibinfo {author} {\bibfnamefont {N.}~\bibnamefont
  {Sakai}},\ }\href {\doibase 10.1007/BF01573998} {\bibfield  {journal}
  {\bibinfo  {journal} {Z.Phys.}\ }\textbf {\bibinfo {volume} {C11}},\ \bibinfo
  {pages} {153} (\bibinfo {year} {1981})}\BibitemShut {NoStop}%
%%CITATION = ZEPYA,C11,153;%%
\bibitem [{\citenamefont {Dimopoulos}\ and\ \citenamefont
  {Georgi}(1981)}]{Dimopoulos:1981zb}%
  \BibitemOpen
  \bibfield  {author} {\bibinfo {author} {\bibfnamefont {S.}~\bibnamefont
  {Dimopoulos}}\ and\ \bibinfo {author} {\bibfnamefont {H.}~\bibnamefont
  {Georgi}},\ }\href {\doibase 10.1016/0550-3213(81)90522-8} {\bibfield
  {journal} {\bibinfo  {journal} {Nucl.Phys.}\ }\textbf {\bibinfo {volume}
  {B193}},\ \bibinfo {pages} {150} (\bibinfo {year} {1981})}\BibitemShut
  {NoStop}%
%%CITATION = NUPHA,B193,150;%%
\bibitem [{\citenamefont {Kawasaki}\ \emph {et~al.}(2008)\citenamefont
  {Kawasaki}, \citenamefont {Kohri}, \citenamefont {Moroi},\ and\ \citenamefont
  {Yotsuyanagi}}]{Kawasaki:2008qe}%
  \BibitemOpen
  \bibfield  {author} {\bibinfo {author} {\bibfnamefont {M.}~\bibnamefont
  {Kawasaki}}, \bibinfo {author} {\bibfnamefont {K.}~\bibnamefont {Kohri}},
  \bibinfo {author} {\bibfnamefont {T.}~\bibnamefont {Moroi}}, \ and\ \bibinfo
  {author} {\bibfnamefont {A.}~\bibnamefont {Yotsuyanagi}},\ }\href {\doibase
  10.1103/PhysRevD.78.065011} {\bibfield  {journal} {\bibinfo  {journal}
  {Phys.Rev.}\ }\textbf {\bibinfo {volume} {D78}},\ \bibinfo {pages} {065011}
  (\bibinfo {year} {2008})},\ \Eprint
  {http://arxiv.org/abs/0804.3745}{arXiv:0804.3745 [hep-ph]}\BibitemShut
  {NoStop}%
%%CITATION = ARXIV:0804.3745;%%
\bibitem [{\citenamefont {Wells}(2003)}]{Wells:2003tf}%
  \BibitemOpen
  \bibfield  {author} {\bibinfo {author} {\bibfnamefont {J.~D.}\ \bibnamefont
  {Wells}},\ }\href@noop {} {\  (\bibinfo {year} {2003})},\ \Eprint
  {http://arxiv.org/abs/hep-ph/0306127}{arXiv:hep-ph/0306127
  [hep-ph]}\BibitemShut {NoStop}%
%%CITATION = HEP-PH/0306127;%%
\bibitem [{\citenamefont {Wells}(2005)}]{Wells:2004di}%
  \BibitemOpen
  \bibfield  {author} {\bibinfo {author} {\bibfnamefont {J.~D.}\ \bibnamefont
  {Wells}},\ }\href {\doibase 10.1103/PhysRevD.71.015013} {\bibfield  {journal}
  {\bibinfo  {journal} {Phys.Rev.}\ }\textbf {\bibinfo {volume} {D71}},\
  \bibinfo {pages} {015013} (\bibinfo {year} {2005})},\ \Eprint
  {http://arxiv.org/abs/hep-ph/0411041}{arXiv:hep-ph/0411041
  [hep-ph]}\BibitemShut {NoStop}%
%%CITATION = HEP-PH/0411041;%%
\bibitem [{\citenamefont {Arkani-Hamed}\ and\ \citenamefont
  {Dimopoulos}(2005)}]{ArkaniHamed:2004fb}%
  \BibitemOpen
  \bibfield  {author} {\bibinfo {author} {\bibfnamefont {N.}~\bibnamefont
  {Arkani-Hamed}}\ and\ \bibinfo {author} {\bibfnamefont {S.}~\bibnamefont
  {Dimopoulos}},\ }\href {\doibase 10.1088/1126-6708/2005/06/073} {\bibfield
  {journal} {\bibinfo  {journal} {JHEP}\ }\textbf {\bibinfo {volume} {0506}},\
  \bibinfo {pages} {073} (\bibinfo {year} {2005})},\ \Eprint
  {http://arxiv.org/abs/hep-th/0405159}{arXiv:hep-th/0405159
  [hep-th]}\BibitemShut {NoStop}%
%%CITATION = HEP-TH/0405159;%%
\bibitem [{\citenamefont {Giudice}\ and\ \citenamefont
  {Romanino}(2004)}]{Giudice:2004tc}%
  \BibitemOpen
  \bibfield  {author} {\bibinfo {author} {\bibfnamefont {G.}~\bibnamefont
  {Giudice}}\ and\ \bibinfo {author} {\bibfnamefont {A.}~\bibnamefont
  {Romanino}},\ }\href {\doibase 10.1016/j.nuclphysb.2004.11.048} {\bibfield
  {journal} {\bibinfo  {journal} {Nucl.Phys.}\ }\textbf {\bibinfo {volume}
  {B699}},\ \bibinfo {pages} {65} (\bibinfo {year} {2004})},\ \Eprint
  {http://arxiv.org/abs/hep-ph/0406088}{arXiv:hep-ph/0406088
  [hep-ph]}\BibitemShut {NoStop}%
%%CITATION = HEP-PH/0406088;%%
\bibitem [{\citenamefont {Arkani-Hamed}\ \emph
  {et~al.}(2005{\natexlab{a}})\citenamefont {Arkani-Hamed}, \citenamefont
  {Dimopoulos}, \citenamefont {Giudice},\ and\ \citenamefont
  {Romanino}}]{ArkaniHamed:2004yi}%
  \BibitemOpen
  \bibfield  {author} {\bibinfo {author} {\bibfnamefont {N.}~\bibnamefont
  {Arkani-Hamed}}, \bibinfo {author} {\bibfnamefont {S.}~\bibnamefont
  {Dimopoulos}}, \bibinfo {author} {\bibfnamefont {G.}~\bibnamefont {Giudice}},
  \ and\ \bibinfo {author} {\bibfnamefont {A.}~\bibnamefont {Romanino}},\
  }\href {\doibase 10.1016/j.nuclphysb.2004.12.026} {\bibfield  {journal}
  {\bibinfo  {journal} {Nucl.Phys.}\ }\textbf {\bibinfo {volume} {B709}},\
  \bibinfo {pages} {3} (\bibinfo {year} {2005}{\natexlab{a}})},\ \Eprint
  {http://arxiv.org/abs/hep-ph/0409232}{arXiv:hep-ph/0409232
  [hep-ph]}\BibitemShut {NoStop}%
%%CITATION = HEP-PH/0409232;%%
\bibitem [{\citenamefont {Arkani-Hamed}\ \emph
  {et~al.}(2005{\natexlab{b}})\citenamefont {Arkani-Hamed}, \citenamefont
  {Dimopoulos},\ and\ \citenamefont {Kachru}}]{ArkaniHamed:2005yv}%
  \BibitemOpen
  \bibfield  {author} {\bibinfo {author} {\bibfnamefont {N.}~\bibnamefont
  {Arkani-Hamed}}, \bibinfo {author} {\bibfnamefont {S.}~\bibnamefont
  {Dimopoulos}}, \ and\ \bibinfo {author} {\bibfnamefont {S.}~\bibnamefont
  {Kachru}},\ }\href@noop {} {\  (\bibinfo {year} {2005}{\natexlab{b}})},\
  \Eprint {http://arxiv.org/abs/hep-th/0501082}{arXiv:hep-th/0501082
  [hep-th]}\BibitemShut {NoStop}%
%%CITATION = HEP-TH/0501082;%%
\bibitem [{\citenamefont {Hall}\ and\ \citenamefont
  {Nomura}(2012)}]{Hall:2011jd}%
  \BibitemOpen
  \bibfield  {author} {\bibinfo {author} {\bibfnamefont {L.~J.}\ \bibnamefont
  {Hall}}\ and\ \bibinfo {author} {\bibfnamefont {Y.}~\bibnamefont {Nomura}},\
  }\href {\doibase 10.1007/JHEP01(2012)082} {\bibfield  {journal} {\bibinfo
  {journal} {JHEP}\ }\textbf {\bibinfo {volume} {1201}},\ \bibinfo {pages}
  {082} (\bibinfo {year} {2012})},\ \Eprint
  {http://arxiv.org/abs/1111.4519}{arXiv:1111.4519 [hep-ph]}\BibitemShut
  {NoStop}%
%%CITATION = ARXIV:1111.4519;%%
\bibitem [{\citenamefont {Hall}\ \emph {et~al.}(2013)\citenamefont {Hall},
  \citenamefont {Nomura},\ and\ \citenamefont {Shirai}}]{Hall:2012zp}%
  \BibitemOpen
  \bibfield  {author} {\bibinfo {author} {\bibfnamefont {L.~J.}\ \bibnamefont
  {Hall}}, \bibinfo {author} {\bibfnamefont {Y.}~\bibnamefont {Nomura}}, \ and\
  \bibinfo {author} {\bibfnamefont {S.}~\bibnamefont {Shirai}},\ }\href
  {\doibase 10.1007/JHEP01(2013)036} {\bibfield  {journal} {\bibinfo  {journal}
  {JHEP}\ }\textbf {\bibinfo {volume} {1301}},\ \bibinfo {pages} {036}
  (\bibinfo {year} {2013})},\ \Eprint
  {http://arxiv.org/abs/1210.2395}{arXiv:1210.2395 [hep-ph]}\BibitemShut
  {NoStop}%
%%CITATION = ARXIV:1210.2395;%%
\bibitem [{\citenamefont {Ibe}\ \emph {et~al.}(2007)\citenamefont {Ibe},
  \citenamefont {Moroi},\ and\ \citenamefont {Yanagida}}]{Ibe:2006de}%
  \BibitemOpen
  \bibfield  {author} {\bibinfo {author} {\bibfnamefont {M.}~\bibnamefont
  {Ibe}}, \bibinfo {author} {\bibfnamefont {T.}~\bibnamefont {Moroi}}, \ and\
  \bibinfo {author} {\bibfnamefont {T.}~\bibnamefont {Yanagida}},\ }\href
  {\doibase 10.1016/j.physletb.2006.11.061} {\bibfield  {journal} {\bibinfo
  {journal} {Phys.Lett.}\ }\textbf {\bibinfo {volume} {B644}},\ \bibinfo
  {pages} {355} (\bibinfo {year} {2007})},\ \Eprint
  {http://arxiv.org/abs/hep-ph/0610277}{arXiv:hep-ph/0610277
  [hep-ph]}\BibitemShut {NoStop}%
%%CITATION = HEP-PH/0610277;%%
\bibitem [{\citenamefont {Ibe}\ and\ \citenamefont
  {Yanagida}(2012)}]{Ibe:2011aa}%
  \BibitemOpen
  \bibfield  {author} {\bibinfo {author} {\bibfnamefont {M.}~\bibnamefont
  {Ibe}}\ and\ \bibinfo {author} {\bibfnamefont {T.~T.}\ \bibnamefont
  {Yanagida}},\ }\href {\doibase 10.1016/j.physletb.2012.02.034} {\bibfield
  {journal} {\bibinfo  {journal} {Phys.Lett.}\ }\textbf {\bibinfo {volume}
  {B709}},\ \bibinfo {pages} {374} (\bibinfo {year} {2012})},\ \Eprint
  {http://arxiv.org/abs/1112.2462}{arXiv:1112.2462 [hep-ph]}\BibitemShut
  {NoStop}%
%%CITATION = ARXIV:1112.2462;%%
\bibitem [{\citenamefont {Ibe}\ \emph {et~al.}(2012)\citenamefont {Ibe},
  \citenamefont {Matsumoto},\ and\ \citenamefont {Yanagida}}]{Ibe:2012hu}%
  \BibitemOpen
  \bibfield  {author} {\bibinfo {author} {\bibfnamefont {M.}~\bibnamefont
  {Ibe}}, \bibinfo {author} {\bibfnamefont {S.}~\bibnamefont {Matsumoto}}, \
  and\ \bibinfo {author} {\bibfnamefont {T.~T.}\ \bibnamefont {Yanagida}},\
  }\href {\doibase 10.1103/PhysRevD.85.095011} {\bibfield  {journal} {\bibinfo
  {journal} {Phys.Rev.}\ }\textbf {\bibinfo {volume} {D85}},\ \bibinfo {pages}
  {095011} (\bibinfo {year} {2012})},\ \Eprint
  {http://arxiv.org/abs/1202.2253}{arXiv:1202.2253 [hep-ph]}\BibitemShut
  {NoStop}%
%%CITATION = ARXIV:1202.2253;%%
\bibitem [{\citenamefont {Arvanitaki}\ \emph {et~al.}(2013)\citenamefont
  {Arvanitaki}, \citenamefont {Craig}, \citenamefont {Dimopoulos},\ and\
  \citenamefont {Villadoro}}]{Arvanitaki:2012ps}%
  \BibitemOpen
  \bibfield  {author} {\bibinfo {author} {\bibfnamefont {A.}~\bibnamefont
  {Arvanitaki}}, \bibinfo {author} {\bibfnamefont {N.}~\bibnamefont {Craig}},
  \bibinfo {author} {\bibfnamefont {S.}~\bibnamefont {Dimopoulos}}, \ and\
  \bibinfo {author} {\bibfnamefont {G.}~\bibnamefont {Villadoro}},\ }\href
  {\doibase 10.1007/JHEP02(2013)126} {\bibfield  {journal} {\bibinfo  {journal}
  {JHEP}\ }\textbf {\bibinfo {volume} {1302}},\ \bibinfo {pages} {126}
  (\bibinfo {year} {2013})},\ \Eprint
  {http://arxiv.org/abs/1210.0555}{arXiv:1210.0555 [hep-ph]}\BibitemShut
  {NoStop}%
%%CITATION = ARXIV:1210.0555;%%
\bibitem [{\citenamefont {Arkani-Hamed}\ \emph {et~al.}(2012)\citenamefont
  {Arkani-Hamed}, \citenamefont {Gupta}, \citenamefont {Kaplan}, \citenamefont
  {Weiner},\ and\ \citenamefont {Zorawski}}]{ArkaniHamed:2012gw}%
  \BibitemOpen
  \bibfield  {author} {\bibinfo {author} {\bibfnamefont {N.}~\bibnamefont
  {Arkani-Hamed}}, \bibinfo {author} {\bibfnamefont {A.}~\bibnamefont {Gupta}},
  \bibinfo {author} {\bibfnamefont {D.~E.}\ \bibnamefont {Kaplan}}, \bibinfo
  {author} {\bibfnamefont {N.}~\bibnamefont {Weiner}}, \ and\ \bibinfo {author}
  {\bibfnamefont {T.}~\bibnamefont {Zorawski}},\ }\href@noop {} {\  (\bibinfo
  {year} {2012})},\ \Eprint {http://arxiv.org/abs/1212.6971}{arXiv:1212.6971
  [hep-ph]}\BibitemShut {NoStop}%
%%CITATION = ARXIV:1212.6971;%%
\bibitem [{\citenamefont {Evans}\ \emph
  {et~al.}(2013{\natexlab{a}})\citenamefont {Evans}, \citenamefont {Ibe},
  \citenamefont {Olive},\ and\ \citenamefont {Yanagida}}]{Evans:2013lpa}%
  \BibitemOpen
  \bibfield  {author} {\bibinfo {author} {\bibfnamefont {J.~L.}\ \bibnamefont
  {Evans}}, \bibinfo {author} {\bibfnamefont {M.}~\bibnamefont {Ibe}}, \bibinfo
  {author} {\bibfnamefont {K.~A.}\ \bibnamefont {Olive}}, \ and\ \bibinfo
  {author} {\bibfnamefont {T.~T.}\ \bibnamefont {Yanagida}},\ }\href {\doibase
  10.1140/epjc/s10052-013-2468-9} {\bibfield  {journal} {\bibinfo  {journal}
  {Eur.Phys.J.}\ }\textbf {\bibinfo {volume} {C73}},\ \bibinfo {pages} {2468}
  (\bibinfo {year} {2013}{\natexlab{a}})},\ \Eprint
  {http://arxiv.org/abs/1302.5346}{arXiv:1302.5346 [hep-ph]}\BibitemShut
  {NoStop}%
%%CITATION = ARXIV:1302.5346;%%
\bibitem [{\citenamefont {Evans}\ \emph
  {et~al.}(2013{\natexlab{b}})\citenamefont {Evans}, \citenamefont {Olive},
  \citenamefont {Ibe},\ and\ \citenamefont {Yanagida}}]{Evans:2013dza}%
  \BibitemOpen
  \bibfield  {author} {\bibinfo {author} {\bibfnamefont {J.~L.}\ \bibnamefont
  {Evans}}, \bibinfo {author} {\bibfnamefont {K.~A.}\ \bibnamefont {Olive}},
  \bibinfo {author} {\bibfnamefont {M.}~\bibnamefont {Ibe}}, \ and\ \bibinfo
  {author} {\bibfnamefont {T.~T.}\ \bibnamefont {Yanagida}},\ }\href {\doibase
  10.1140/epjc/s10052-013-2611-7} {\bibfield  {journal} {\bibinfo  {journal}
  {Eur.Phys.J.}\ }\textbf {\bibinfo {volume} {C73}},\ \bibinfo {pages} {2611}
  (\bibinfo {year} {2013}{\natexlab{b}})},\ \Eprint
  {http://arxiv.org/abs/1305.7461}{arXiv:1305.7461 [hep-ph]}\BibitemShut
  {NoStop}%
%%CITATION = ARXIV:1305.7461;%%
\bibitem [{\citenamefont {Evans}\ \emph
  {et~al.}(2014{\natexlab{a}})\citenamefont {Evans}, \citenamefont {Ibe},
  \citenamefont {Olive},\ and\ \citenamefont {Yanagida}}]{Evans:2013uza}%
  \BibitemOpen
  \bibfield  {author} {\bibinfo {author} {\bibfnamefont {J.~L.}\ \bibnamefont
  {Evans}}, \bibinfo {author} {\bibfnamefont {M.}~\bibnamefont {Ibe}}, \bibinfo
  {author} {\bibfnamefont {K.~A.}\ \bibnamefont {Olive}}, \ and\ \bibinfo
  {author} {\bibfnamefont {T.~T.}\ \bibnamefont {Yanagida}},\ }\href {\doibase
  10.1140/epjc/s10052-014-2775-9} {\bibfield  {journal} {\bibinfo  {journal}
  {Eur.Phys.J.}\ }\textbf {\bibinfo {volume} {C74}},\ \bibinfo {pages} {2775}
  (\bibinfo {year} {2014}{\natexlab{a}})},\ \Eprint
  {http://arxiv.org/abs/1312.1984}{arXiv:1312.1984 [hep-ph]}\BibitemShut
  {NoStop}%
%%CITATION = ARXIV:1312.1984;%%
\bibitem [{\citenamefont {Evans}\ \emph
  {et~al.}(2014{\natexlab{b}})\citenamefont {Evans}, \citenamefont {Ibe},
  \citenamefont {Olive},\ and\ \citenamefont {Yanagida}}]{Evans:2014hda}%
  \BibitemOpen
  \bibfield  {author} {\bibinfo {author} {\bibfnamefont {J.~L.}\ \bibnamefont
  {Evans}}, \bibinfo {author} {\bibfnamefont {M.}~\bibnamefont {Ibe}}, \bibinfo
  {author} {\bibfnamefont {K.~A.}\ \bibnamefont {Olive}}, \ and\ \bibinfo
  {author} {\bibfnamefont {T.~T.}\ \bibnamefont {Yanagida}},\ }\href {\doibase
  10.1140/epjc/s10052-014-2931-2} {\bibfield  {journal} {\bibinfo  {journal}
  {Eur.Phys.J.}\ }\textbf {\bibinfo {volume} {C74}},\ \bibinfo {pages} {2931}
  (\bibinfo {year} {2014}{\natexlab{b}})},\ \Eprint
  {http://arxiv.org/abs/1402.5989}{arXiv:1402.5989 [hep-ph]}\BibitemShut
  {NoStop}%
%%CITATION = ARXIV:1402.5989;%%
\bibitem [{\citenamefont {Nomura}\ and\ \citenamefont
  {Shirai}(2014)}]{Nomura:2014asa}%
  \BibitemOpen
  \bibfield  {author} {\bibinfo {author} {\bibfnamefont {Y.}~\bibnamefont
  {Nomura}}\ and\ \bibinfo {author} {\bibfnamefont {S.}~\bibnamefont
  {Shirai}},\ }\href {\doibase 10.1103/PhysRevLett.113.111801} {\bibfield
  {journal} {\bibinfo  {journal} {Phys.Rev.Lett.}\ }\textbf {\bibinfo {volume}
  {113}},\ \bibinfo {pages} {111801} (\bibinfo {year} {2014})},\ \Eprint
  {http://arxiv.org/abs/1407.3785}{arXiv:1407.3785 [hep-ph]}\BibitemShut
  {NoStop}%
%%CITATION = ARXIV:1407.3785;%%
\bibitem [{\citenamefont {Giudice}\ \emph {et~al.}(1998)\citenamefont
  {Giudice}, \citenamefont {Luty}, \citenamefont {Murayama},\ and\
  \citenamefont {Rattazzi}}]{Giudice:1998xp}%
  \BibitemOpen
  \bibfield  {author} {\bibinfo {author} {\bibfnamefont {G.~F.}\ \bibnamefont
  {Giudice}}, \bibinfo {author} {\bibfnamefont {M.~A.}\ \bibnamefont {Luty}},
  \bibinfo {author} {\bibfnamefont {H.}~\bibnamefont {Murayama}}, \ and\
  \bibinfo {author} {\bibfnamefont {R.}~\bibnamefont {Rattazzi}},\ }\href
  {\doibase 10.1088/1126-6708/1998/12/027} {\bibfield  {journal} {\bibinfo
  {journal} {JHEP}\ }\textbf {\bibinfo {volume} {9812}},\ \bibinfo {pages}
  {027} (\bibinfo {year} {1998})},\ \Eprint
  {http://arxiv.org/abs/hep-ph/9810442}{arXiv:hep-ph/9810442
  [hep-ph]}\BibitemShut {NoStop}%
%%CITATION = HEP-PH/9810442;%%
\bibitem [{\citenamefont {Randall}\ and\ \citenamefont
  {Sundrum}(1999)}]{Randall:1998uk}%
  \BibitemOpen
  \bibfield  {author} {\bibinfo {author} {\bibfnamefont {L.}~\bibnamefont
  {Randall}}\ and\ \bibinfo {author} {\bibfnamefont {R.}~\bibnamefont
  {Sundrum}},\ }\href {\doibase 10.1016/S0550-3213(99)00359-4} {\bibfield
  {journal} {\bibinfo  {journal} {Nucl.Phys.}\ }\textbf {\bibinfo {volume}
  {B557}},\ \bibinfo {pages} {79} (\bibinfo {year} {1999})},\ \Eprint
  {http://arxiv.org/abs/hep-th/9810155}{arXiv:hep-th/9810155
  [hep-th]}\BibitemShut {NoStop}%
%%CITATION = HEP-TH/9810155;%%
\bibitem [{\citenamefont {Pierce}\ \emph {et~al.}(1997)\citenamefont {Pierce},
  \citenamefont {Bagger}, \citenamefont {Matchev},\ and\ \citenamefont
  {Zhang}}]{Pierce:1996zz}%
  \BibitemOpen
  \bibfield  {author} {\bibinfo {author} {\bibfnamefont {D.~M.}\ \bibnamefont
  {Pierce}}, \bibinfo {author} {\bibfnamefont {J.~A.}\ \bibnamefont {Bagger}},
  \bibinfo {author} {\bibfnamefont {K.~T.}\ \bibnamefont {Matchev}}, \ and\
  \bibinfo {author} {\bibfnamefont {R.-j.}\ \bibnamefont {Zhang}},\ }\href
  {\doibase 10.1016/S0550-3213(96)00683-9} {\bibfield  {journal} {\bibinfo
  {journal} {Nucl.Phys.}\ }\textbf {\bibinfo {volume} {B491}},\ \bibinfo
  {pages} {3} (\bibinfo {year} {1997})},\ \Eprint
  {http://arxiv.org/abs/hep-ph/9606211}{arXiv:hep-ph/9606211
  [hep-ph]}\BibitemShut {NoStop}%
%%CITATION = HEP-PH/9606211;%%
\bibitem [{\citenamefont {Pomarol}\ and\ \citenamefont
  {Rattazzi}(1999)}]{Pomarol:1999ie}%
  \BibitemOpen
  \bibfield  {author} {\bibinfo {author} {\bibfnamefont {A.}~\bibnamefont
  {Pomarol}}\ and\ \bibinfo {author} {\bibfnamefont {R.}~\bibnamefont
  {Rattazzi}},\ }\href {\doibase 10.1088/1126-6708/1999/05/013} {\bibfield
  {journal} {\bibinfo  {journal} {JHEP}\ }\textbf {\bibinfo {volume} {9905}},\
  \bibinfo {pages} {013} (\bibinfo {year} {1999})},\ \Eprint
  {http://arxiv.org/abs/hep-ph/9903448}{arXiv:hep-ph/9903448
  [hep-ph]}\BibitemShut {NoStop}%
%%CITATION = HEP-PH/9903448;%%
\bibitem [{\citenamefont {Nelson}\ and\ \citenamefont
  {Weiner}(2002)}]{Nelson:2002sa}%
  \BibitemOpen
  \bibfield  {author} {\bibinfo {author} {\bibfnamefont {A.~E.}\ \bibnamefont
  {Nelson}}\ and\ \bibinfo {author} {\bibfnamefont {N.~J.}\ \bibnamefont
  {Weiner}},\ }\href@noop {} {\  (\bibinfo {year} {2002})},\ \Eprint
  {http://arxiv.org/abs/hep-ph/0210288}{arXiv:hep-ph/0210288
  [hep-ph]}\BibitemShut {NoStop}%
%%CITATION = HEP-PH/0210288;%%
\bibitem [{\citenamefont {Hsieh}\ and\ \citenamefont
  {Luty}(2007)}]{Hsieh:2006ig}%
  \BibitemOpen
  \bibfield  {author} {\bibinfo {author} {\bibfnamefont {K.}~\bibnamefont
  {Hsieh}}\ and\ \bibinfo {author} {\bibfnamefont {M.~A.}\ \bibnamefont
  {Luty}},\ }\href {\doibase 10.1088/1126-6708/2007/06/062} {\bibfield
  {journal} {\bibinfo  {journal} {JHEP}\ }\textbf {\bibinfo {volume} {0706}},\
  \bibinfo {pages} {062} (\bibinfo {year} {2007})},\ \Eprint
  {http://arxiv.org/abs/hep-ph/0604256}{arXiv:hep-ph/0604256
  [hep-ph]}\BibitemShut {NoStop}%
%%CITATION = HEP-PH/0604256;%%
\bibitem [{\citenamefont {Gupta}\ \emph {et~al.}(2013)\citenamefont {Gupta},
  \citenamefont {Kaplan},\ and\ \citenamefont {Zorawski}}]{Gupta:2012gu}%
  \BibitemOpen
  \bibfield  {author} {\bibinfo {author} {\bibfnamefont {A.}~\bibnamefont
  {Gupta}}, \bibinfo {author} {\bibfnamefont {D.~E.}\ \bibnamefont {Kaplan}}, \
  and\ \bibinfo {author} {\bibfnamefont {T.}~\bibnamefont {Zorawski}},\ }\href
  {\doibase 10.1007/JHEP11(2013)149} {\bibfield  {journal} {\bibinfo  {journal}
  {JHEP}\ }\textbf {\bibinfo {volume} {1311}},\ \bibinfo {pages} {149}
  (\bibinfo {year} {2013})},\ \Eprint
  {http://arxiv.org/abs/1212.6969}{arXiv:1212.6969 [hep-ph]}\BibitemShut
  {NoStop}%
%%CITATION = ARXIV:1212.6969;%%
\bibitem [{\citenamefont {Nakayama}\ and\ \citenamefont
  {Yanagida}(2013)}]{Nakayama:2013uta}%
  \BibitemOpen
  \bibfield  {author} {\bibinfo {author} {\bibfnamefont {K.}~\bibnamefont
  {Nakayama}}\ and\ \bibinfo {author} {\bibfnamefont {T.~T.}\ \bibnamefont
  {Yanagida}},\ }\href {\doibase 10.1016/j.physletb.2013.04.002} {\bibfield
  {journal} {\bibinfo  {journal} {Phys.Lett.}\ }\textbf {\bibinfo {volume}
  {B722}},\ \bibinfo {pages} {107} (\bibinfo {year} {2013})},\ \Eprint
  {http://arxiv.org/abs/1302.3332}{arXiv:1302.3332 [hep-ph]}\BibitemShut
  {NoStop}%
%%CITATION = ARXIV:1302.3332;%%
\bibitem [{\citenamefont {Harigaya}\ \emph {et~al.}(2013)\citenamefont
  {Harigaya}, \citenamefont {Ibe},\ and\ \citenamefont
  {Yanagida}}]{Harigaya:2013asa}%
  \BibitemOpen
  \bibfield  {author} {\bibinfo {author} {\bibfnamefont {K.}~\bibnamefont
  {Harigaya}}, \bibinfo {author} {\bibfnamefont {M.}~\bibnamefont {Ibe}}, \
  and\ \bibinfo {author} {\bibfnamefont {T.~T.}\ \bibnamefont {Yanagida}},\
  }\href {\doibase 10.1007/JHEP12(2013)016} {\bibfield  {journal} {\bibinfo
  {journal} {JHEP}\ }\textbf {\bibinfo {volume} {1312}},\ \bibinfo {pages}
  {016} (\bibinfo {year} {2013})},\ \Eprint
  {http://arxiv.org/abs/1310.0643}{arXiv:1310.0643 [hep-ph]}\BibitemShut
  {NoStop}%
%%CITATION = ARXIV:1310.0643;%%
\bibitem [{\citenamefont {Evans}\ and\ \citenamefont
  {Olive}(2014)}]{Evans:2014xpa}%
  \BibitemOpen
  \bibfield  {author} {\bibinfo {author} {\bibfnamefont {J.~L.}\ \bibnamefont
  {Evans}}\ and\ \bibinfo {author} {\bibfnamefont {K.~A.}\ \bibnamefont
  {Olive}},\ }\href {\doibase 10.1103/PhysRevD.90.115020} {\bibfield  {journal}
  {\bibinfo  {journal} {Phys.Rev.}\ }\textbf {\bibinfo {volume} {D90}},\
  \bibinfo {pages} {115020} (\bibinfo {year} {2014})},\ \Eprint
  {http://arxiv.org/abs/1408.5102}{arXiv:1408.5102 [hep-ph]}\BibitemShut
  {NoStop}%
%%CITATION = ARXIV:1408.5102;%%
\bibitem [{\citenamefont {Peccei}\ and\ \citenamefont
  {Quinn}(1977)}]{Peccei:1977hh}%
  \BibitemOpen
  \bibfield  {author} {\bibinfo {author} {\bibfnamefont {R.}~\bibnamefont
  {Peccei}}\ and\ \bibinfo {author} {\bibfnamefont {H.~R.}\ \bibnamefont
  {Quinn}},\ }\href {\doibase 10.1103/PhysRevLett.38.1440} {\bibfield
  {journal} {\bibinfo  {journal} {Phys.Rev.Lett.}\ }\textbf {\bibinfo {volume}
  {38}},\ \bibinfo {pages} {1440} (\bibinfo {year} {1977})}\BibitemShut
  {NoStop}%
%%CITATION = PRLTA,38,1440;%%
\bibitem [{\citenamefont {Bae}\ \emph {et~al.}(2015)\citenamefont {Bae},
  \citenamefont {Baer},\ and\ \citenamefont {Serce}}]{Bae:2014yta}%
  \BibitemOpen
  \bibfield  {author} {\bibinfo {author} {\bibfnamefont {K.~J.}\ \bibnamefont
  {Bae}}, \bibinfo {author} {\bibfnamefont {H.}~\bibnamefont {Baer}}, \ and\
  \bibinfo {author} {\bibfnamefont {H.}~\bibnamefont {Serce}},\ }\href
  {\doibase 10.1103/PhysRevD.91.015003} {\bibfield  {journal} {\bibinfo
  {journal} {Phys.Rev.}\ }\textbf {\bibinfo {volume} {D91}},\ \bibinfo {pages}
  {015003} (\bibinfo {year} {2015})},\ \Eprint
  {http://arxiv.org/abs/1410.7500}{arXiv:1410.7500 [hep-ph]}\BibitemShut
  {NoStop}%
%%CITATION = ARXIV:1410.7500;%%
\bibitem [{\citenamefont {Evans}\ \emph
  {et~al.}(2015{\natexlab{b}})\citenamefont {Evans}, \citenamefont {Ibe},
  \citenamefont {Olive},\ and\ \citenamefont {Yanagida}}]{Evans:2014pxa}%
  \BibitemOpen
  \bibfield  {author} {\bibinfo {author} {\bibfnamefont {J.~L.}\ \bibnamefont
  {Evans}}, \bibinfo {author} {\bibfnamefont {M.}~\bibnamefont {Ibe}}, \bibinfo
  {author} {\bibfnamefont {K.~A.}\ \bibnamefont {Olive}}, \ and\ \bibinfo
  {author} {\bibfnamefont {T.~T.}\ \bibnamefont {Yanagida}},\ }\href {\doibase
  10.1103/PhysRevD.91.055008} {\bibfield  {journal} {\bibinfo  {journal}
  {Phys.Rev.}\ }\textbf {\bibinfo {volume} {D91}},\ \bibinfo {pages} {055008}
  (\bibinfo {year} {2015}{\natexlab{b}})},\ \Eprint
  {http://arxiv.org/abs/1412.3403}{arXiv:1412.3403 [hep-ph]}\BibitemShut
  {NoStop}%
%%CITATION = ARXIV:1412.3403;%%
\bibitem [{\citenamefont {Hisano}\ \emph
  {et~al.}(2013{\natexlab{b}})\citenamefont {Hisano}, \citenamefont
  {Kuwahara},\ and\ \citenamefont {Nagata}}]{Hisano:2013cqa}%
  \BibitemOpen
  \bibfield  {author} {\bibinfo {author} {\bibfnamefont {J.}~\bibnamefont
  {Hisano}}, \bibinfo {author} {\bibfnamefont {T.}~\bibnamefont {Kuwahara}}, \
  and\ \bibinfo {author} {\bibfnamefont {N.}~\bibnamefont {Nagata}},\ }\href
  {\doibase 10.1016/j.physletb.2013.05.017} {\bibfield  {journal} {\bibinfo
  {journal} {Phys.Lett.}\ }\textbf {\bibinfo {volume} {B723}},\ \bibinfo
  {pages} {324} (\bibinfo {year} {2013}{\natexlab{b}})},\ \Eprint
  {http://arxiv.org/abs/1304.0343}{arXiv:1304.0343 [hep-ph]}\BibitemShut
  {NoStop}%
%%CITATION = ARXIV:1304.0343;%%
\bibitem [{\citenamefont {Hisano}\ \emph {et~al.}(2007)\citenamefont {Hisano},
  \citenamefont {Matsumoto}, \citenamefont {Nagai}, \citenamefont {Saito},\
  and\ \citenamefont {Senami}}]{Hisano:2006nn}%
  \BibitemOpen
  \bibfield  {author} {\bibinfo {author} {\bibfnamefont {J.}~\bibnamefont
  {Hisano}}, \bibinfo {author} {\bibfnamefont {S.}~\bibnamefont {Matsumoto}},
  \bibinfo {author} {\bibfnamefont {M.}~\bibnamefont {Nagai}}, \bibinfo
  {author} {\bibfnamefont {O.}~\bibnamefont {Saito}}, \ and\ \bibinfo {author}
  {\bibfnamefont {M.}~\bibnamefont {Senami}},\ }\href {\doibase
  10.1016/j.physletb.2007.01.012} {\bibfield  {journal} {\bibinfo  {journal}
  {Phys.Lett.}\ }\textbf {\bibinfo {volume} {B646}},\ \bibinfo {pages} {34}
  (\bibinfo {year} {2007})},\ \Eprint
  {http://arxiv.org/abs/hep-ph/0610249}{arXiv:hep-ph/0610249
  [hep-ph]}\BibitemShut {NoStop}%
%%CITATION = HEP-PH/0610249;%%
\bibitem [{\citenamefont {Aad}\ \emph {et~al.}(2013{\natexlab{a}})\citenamefont
  {Aad} \emph {et~al.}}]{Aad:2013yna}%
  \BibitemOpen
  \bibfield  {author} {\bibinfo {author} {\bibfnamefont {G.}~\bibnamefont
  {Aad}} \emph {et~al.} (\bibinfo {collaboration} {ATLAS}),\ }\href {\doibase
  10.1103/PhysRevD.88.112006} {\bibfield  {journal} {\bibinfo  {journal}
  {Phys.Rev.}\ }\textbf {\bibinfo {volume} {D88}},\ \bibinfo {pages} {112006}
  (\bibinfo {year} {2013}{\natexlab{a}})},\ \Eprint
  {http://arxiv.org/abs/1310.3675}{arXiv:1310.3675 [hep-ex]}\BibitemShut
  {NoStop}%
%%CITATION = ARXIV:1310.3675;%%
\bibitem [{\citenamefont {Cohen}\ \emph {et~al.}(2013)\citenamefont {Cohen},
  \citenamefont {Lisanti}, \citenamefont {Pierce},\ and\ \citenamefont
  {Slatyer}}]{Cohen:2013ama}%
  \BibitemOpen
  \bibfield  {author} {\bibinfo {author} {\bibfnamefont {T.}~\bibnamefont
  {Cohen}}, \bibinfo {author} {\bibfnamefont {M.}~\bibnamefont {Lisanti}},
  \bibinfo {author} {\bibfnamefont {A.}~\bibnamefont {Pierce}}, \ and\ \bibinfo
  {author} {\bibfnamefont {T.~R.}\ \bibnamefont {Slatyer}},\ }\href {\doibase
  10.1088/1475-7516/2013/10/061} {\bibfield  {journal} {\bibinfo  {journal}
  {JCAP}\ }\textbf {\bibinfo {volume} {1310}},\ \bibinfo {pages} {061}
  (\bibinfo {year} {2013})},\ \Eprint
  {http://arxiv.org/abs/1307.4082}{arXiv:1307.4082}\BibitemShut {NoStop}%
%%CITATION = ARXIV:1307.4082;%%
\bibitem [{\citenamefont {Fan}\ and\ \citenamefont
  {Reece}(2013)}]{Fan:2013faa}%
  \BibitemOpen
  \bibfield  {author} {\bibinfo {author} {\bibfnamefont {J.}~\bibnamefont
  {Fan}}\ and\ \bibinfo {author} {\bibfnamefont {M.}~\bibnamefont {Reece}},\
  }\href {\doibase 10.1007/JHEP10(2013)124} {\bibfield  {journal} {\bibinfo
  {journal} {JHEP}\ }\textbf {\bibinfo {volume} {1310}},\ \bibinfo {pages}
  {124} (\bibinfo {year} {2013})},\ \Eprint
  {http://arxiv.org/abs/1307.4400}{arXiv:1307.4400 [hep-ph]}\BibitemShut
  {NoStop}%
%%CITATION = ARXIV:1307.4400;%%
\bibitem [{\citenamefont {Hryczuk}\ \emph {et~al.}(2014)\citenamefont
  {Hryczuk}, \citenamefont {Cholis}, \citenamefont {Iengo}, \citenamefont
  {Tavakoli},\ and\ \citenamefont {Ullio}}]{Hryczuk:2014hpa}%
  \BibitemOpen
  \bibfield  {author} {\bibinfo {author} {\bibfnamefont {A.}~\bibnamefont
  {Hryczuk}}, \bibinfo {author} {\bibfnamefont {I.}~\bibnamefont {Cholis}},
  \bibinfo {author} {\bibfnamefont {R.}~\bibnamefont {Iengo}}, \bibinfo
  {author} {\bibfnamefont {M.}~\bibnamefont {Tavakoli}}, \ and\ \bibinfo
  {author} {\bibfnamefont {P.}~\bibnamefont {Ullio}},\ }\href {\doibase
  10.1088/1475-7516/2014/07/031} {\bibfield  {journal} {\bibinfo  {journal}
  {JCAP}\ }\textbf {\bibinfo {volume} {1407}},\ \bibinfo {pages} {031}
  (\bibinfo {year} {2014})},\ \Eprint
  {http://arxiv.org/abs/1401.6212}{arXiv:1401.6212 [astro-ph.HE]}\BibitemShut
  {NoStop}%
%%CITATION = ARXIV:1401.6212;%%
\bibitem [{\citenamefont {Bhattacherjee}\ \emph
  {et~al.}(2014{\natexlab{a}})\citenamefont {Bhattacherjee}, \citenamefont
  {Ibe}, \citenamefont {Ichikawa}, \citenamefont {Matsumoto},\ and\
  \citenamefont {Nishiyama}}]{Bhattacherjee:2014dya}%
  \BibitemOpen
  \bibfield  {author} {\bibinfo {author} {\bibfnamefont {B.}~\bibnamefont
  {Bhattacherjee}}, \bibinfo {author} {\bibfnamefont {M.}~\bibnamefont {Ibe}},
  \bibinfo {author} {\bibfnamefont {K.}~\bibnamefont {Ichikawa}}, \bibinfo
  {author} {\bibfnamefont {S.}~\bibnamefont {Matsumoto}}, \ and\ \bibinfo
  {author} {\bibfnamefont {K.}~\bibnamefont {Nishiyama}},\ }\href {\doibase
  10.1007/JHEP07(2014)080} {\bibfield  {journal} {\bibinfo  {journal} {JHEP}\
  }\textbf {\bibinfo {volume} {1407}},\ \bibinfo {pages} {080} (\bibinfo {year}
  {2014}{\natexlab{a}})},\ \Eprint
  {http://arxiv.org/abs/1405.4914}{arXiv:1405.4914 [hep-ph]}\BibitemShut
  {NoStop}%
%%CITATION = ARXIV:1405.4914;%%
\bibitem [{\citenamefont {Hisano}\ \emph
  {et~al.}(2010{\natexlab{a}})\citenamefont {Hisano}, \citenamefont
  {Ishiwata},\ and\ \citenamefont {Nagata}}]{Hisano:2010fy}%
  \BibitemOpen
  \bibfield  {author} {\bibinfo {author} {\bibfnamefont {J.}~\bibnamefont
  {Hisano}}, \bibinfo {author} {\bibfnamefont {K.}~\bibnamefont {Ishiwata}}, \
  and\ \bibinfo {author} {\bibfnamefont {N.}~\bibnamefont {Nagata}},\ }\href
  {\doibase 10.1016/j.physletb.2010.05.047} {\bibfield  {journal} {\bibinfo
  {journal} {Phys.Lett.}\ }\textbf {\bibinfo {volume} {B690}},\ \bibinfo
  {pages} {311} (\bibinfo {year} {2010}{\natexlab{a}})},\ \Eprint
  {http://arxiv.org/abs/1004.4090}{arXiv:1004.4090 [hep-ph]}\BibitemShut
  {NoStop}%
%%CITATION = ARXIV:1004.4090;%%
\bibitem [{\citenamefont {Hisano}\ \emph
  {et~al.}(2010{\natexlab{b}})\citenamefont {Hisano}, \citenamefont
  {Ishiwata},\ and\ \citenamefont {Nagata}}]{Hisano:2010ct}%
  \BibitemOpen
  \bibfield  {author} {\bibinfo {author} {\bibfnamefont {J.}~\bibnamefont
  {Hisano}}, \bibinfo {author} {\bibfnamefont {K.}~\bibnamefont {Ishiwata}}, \
  and\ \bibinfo {author} {\bibfnamefont {N.}~\bibnamefont {Nagata}},\ }\href
  {\doibase 10.1103/PhysRevD.82.115007} {\bibfield  {journal} {\bibinfo
  {journal} {Phys.Rev.}\ }\textbf {\bibinfo {volume} {D82}},\ \bibinfo {pages}
  {115007} (\bibinfo {year} {2010}{\natexlab{b}})},\ \Eprint
  {http://arxiv.org/abs/1007.2601}{arXiv:1007.2601 [hep-ph]}\BibitemShut
  {NoStop}%
%%CITATION = ARXIV:1007.2601;%%
\bibitem [{\citenamefont {Hisano}\ \emph {et~al.}(2011)\citenamefont {Hisano},
  \citenamefont {Ishiwata}, \citenamefont {Nagata},\ and\ \citenamefont
  {Takesako}}]{Hisano:2011cs}%
  \BibitemOpen
  \bibfield  {author} {\bibinfo {author} {\bibfnamefont {J.}~\bibnamefont
  {Hisano}}, \bibinfo {author} {\bibfnamefont {K.}~\bibnamefont {Ishiwata}},
  \bibinfo {author} {\bibfnamefont {N.}~\bibnamefont {Nagata}}, \ and\ \bibinfo
  {author} {\bibfnamefont {T.}~\bibnamefont {Takesako}},\ }\href {\doibase
  10.1007/JHEP07(2011)005} {\bibfield  {journal} {\bibinfo  {journal} {JHEP}\
  }\textbf {\bibinfo {volume} {1107}},\ \bibinfo {pages} {005} (\bibinfo {year}
  {2011})},\ \Eprint {http://arxiv.org/abs/1104.0228}{arXiv:1104.0228
  [hep-ph]}\BibitemShut {NoStop}%
%%CITATION = ARXIV:1104.0228;%%
\bibitem [{\citenamefont {Hisano}\ \emph
  {et~al.}(2013{\natexlab{c}})\citenamefont {Hisano}, \citenamefont
  {Ishiwata},\ and\ \citenamefont {Nagata}}]{Hisano:2012wm}%
  \BibitemOpen
  \bibfield  {author} {\bibinfo {author} {\bibfnamefont {J.}~\bibnamefont
  {Hisano}}, \bibinfo {author} {\bibfnamefont {K.}~\bibnamefont {Ishiwata}}, \
  and\ \bibinfo {author} {\bibfnamefont {N.}~\bibnamefont {Nagata}},\ }\href
  {\doibase 10.1103/PhysRevD.87.035020} {\bibfield  {journal} {\bibinfo
  {journal} {Phys.Rev.}\ }\textbf {\bibinfo {volume} {D87}},\ \bibinfo {pages}
  {035020} (\bibinfo {year} {2013}{\natexlab{c}})},\ \Eprint
  {http://arxiv.org/abs/1210.5985}{arXiv:1210.5985 [hep-ph]}\BibitemShut
  {NoStop}%
%%CITATION = ARXIV:1210.5985;%%
\bibitem [{\citenamefont {Hisano}\ \emph {et~al.}(2015)\citenamefont {Hisano},
  \citenamefont {Ishiwata},\ and\ \citenamefont {Nagata}}]{Hisano:2015rsa}%
  \BibitemOpen
  \bibfield  {author} {\bibinfo {author} {\bibfnamefont {J.}~\bibnamefont
  {Hisano}}, \bibinfo {author} {\bibfnamefont {K.}~\bibnamefont {Ishiwata}}, \
  and\ \bibinfo {author} {\bibfnamefont {N.}~\bibnamefont {Nagata}},\ }\href
  {\doibase 10.1007/JHEP06(2015)097} {\bibfield  {journal} {\bibinfo  {journal}
  {JHEP}\ }\textbf {\bibinfo {volume} {1506}},\ \bibinfo {pages} {097}
  (\bibinfo {year} {2015})},\ \Eprint
  {http://arxiv.org/abs/1504.00915}{arXiv:1504.00915 [hep-ph]}\BibitemShut
  {NoStop}%
%%CITATION = ARXIV:1504.00915;%%
\bibitem [{\citenamefont {Cirelli}\ \emph {et~al.}(2007)\citenamefont
  {Cirelli}, \citenamefont {Strumia},\ and\ \citenamefont
  {Tamburini}}]{Cirelli:2007xd}%
  \BibitemOpen
  \bibfield  {author} {\bibinfo {author} {\bibfnamefont {M.}~\bibnamefont
  {Cirelli}}, \bibinfo {author} {\bibfnamefont {A.}~\bibnamefont {Strumia}}, \
  and\ \bibinfo {author} {\bibfnamefont {M.}~\bibnamefont {Tamburini}},\ }\href
  {\doibase 10.1016/j.nuclphysb.2007.07.023} {\bibfield  {journal} {\bibinfo
  {journal} {Nucl.Phys.}\ }\textbf {\bibinfo {volume} {B787}},\ \bibinfo
  {pages} {152} (\bibinfo {year} {2007})},\ \Eprint
  {http://arxiv.org/abs/0706.4071}{arXiv:0706.4071 [hep-ph]}\BibitemShut
  {NoStop}%
%%CITATION = ARXIV:0706.4071;%%
\bibitem [{\citenamefont {Nagata}\ and\ \citenamefont
  {Shirai}(2015)}]{Nagata:2014wma}%
  \BibitemOpen
  \bibfield  {author} {\bibinfo {author} {\bibfnamefont {N.}~\bibnamefont
  {Nagata}}\ and\ \bibinfo {author} {\bibfnamefont {S.}~\bibnamefont
  {Shirai}},\ }\href {\doibase 10.1007/JHEP01(2015)029} {\bibfield  {journal}
  {\bibinfo  {journal} {JHEP}\ }\textbf {\bibinfo {volume} {1501}},\ \bibinfo
  {pages} {029} (\bibinfo {year} {2015})},\ \Eprint
  {http://arxiv.org/abs/1410.4549}{arXiv:1410.4549 [hep-ph]}\BibitemShut
  {NoStop}%
%%CITATION = ARXIV:1410.4549;%%
\bibitem [{\citenamefont {Griest}\ and\ \citenamefont
  {Seckel}(1991)}]{Griest:1990kh}%
  \BibitemOpen
  \bibfield  {author} {\bibinfo {author} {\bibfnamefont {K.}~\bibnamefont
  {Griest}}\ and\ \bibinfo {author} {\bibfnamefont {D.}~\bibnamefont
  {Seckel}},\ }\href {\doibase 10.1103/PhysRevD.43.3191} {\bibfield  {journal}
  {\bibinfo  {journal} {Phys.Rev.}\ }\textbf {\bibinfo {volume} {D43}},\
  \bibinfo {pages} {3191} (\bibinfo {year} {1991})}\BibitemShut {NoStop}%
%%CITATION = PHRVA,D43,3191;%%
\bibitem [{\citenamefont {Baer}\ \emph {et~al.}(2005)\citenamefont {Baer},
  \citenamefont {Krupovnickas}, \citenamefont {Mustafayev}, \citenamefont
  {Park}, \citenamefont {Profumo} \emph {et~al.}}]{Baer:2005jq}%
  \BibitemOpen
  \bibfield  {author} {\bibinfo {author} {\bibfnamefont {H.}~\bibnamefont
  {Baer}}, \bibinfo {author} {\bibfnamefont {T.}~\bibnamefont {Krupovnickas}},
  \bibinfo {author} {\bibfnamefont {A.}~\bibnamefont {Mustafayev}}, \bibinfo
  {author} {\bibfnamefont {E.-K.}\ \bibnamefont {Park}}, \bibinfo {author}
  {\bibfnamefont {S.}~\bibnamefont {Profumo}},  \emph {et~al.},\ }\href
  {\doibase 10.1088/1126-6708/2005/12/011} {\bibfield  {journal} {\bibinfo
  {journal} {JHEP}\ }\textbf {\bibinfo {volume} {0512}},\ \bibinfo {pages}
  {011} (\bibinfo {year} {2005})},\ \Eprint
  {http://arxiv.org/abs/hep-ph/0511034}{arXiv:hep-ph/0511034
  [hep-ph]}\BibitemShut {NoStop}%
%%CITATION = HEP-PH/0511034;%%
\bibitem [{\citenamefont {Profumo}\ and\ \citenamefont
  {Yaguna}(2004)}]{Profumo:2004wk}%
  \BibitemOpen
  \bibfield  {author} {\bibinfo {author} {\bibfnamefont {S.}~\bibnamefont
  {Profumo}}\ and\ \bibinfo {author} {\bibfnamefont {C.}~\bibnamefont
  {Yaguna}},\ }\href {\doibase 10.1103/PhysRevD.69.115009} {\bibfield
  {journal} {\bibinfo  {journal} {Phys.Rev.}\ }\textbf {\bibinfo {volume}
  {D69}},\ \bibinfo {pages} {115009} (\bibinfo {year} {2004})},\ \Eprint
  {http://arxiv.org/abs/hep-ph/0402208}{arXiv:hep-ph/0402208
  [hep-ph]}\BibitemShut {NoStop}%
%%CITATION = HEP-PH/0402208;%%
\bibitem [{\citenamefont {Feldman}\ \emph {et~al.}(2009)\citenamefont
  {Feldman}, \citenamefont {Liu},\ and\ \citenamefont {Nath}}]{Feldman:2009zc}%
  \BibitemOpen
  \bibfield  {author} {\bibinfo {author} {\bibfnamefont {D.}~\bibnamefont
  {Feldman}}, \bibinfo {author} {\bibfnamefont {Z.}~\bibnamefont {Liu}}, \ and\
  \bibinfo {author} {\bibfnamefont {P.}~\bibnamefont {Nath}},\ }\href {\doibase
  10.1103/PhysRevD.80.015007} {\bibfield  {journal} {\bibinfo  {journal}
  {Phys.Rev.}\ }\textbf {\bibinfo {volume} {D80}},\ \bibinfo {pages} {015007}
  (\bibinfo {year} {2009})},\ \Eprint
  {http://arxiv.org/abs/0905.1148}{arXiv:0905.1148 [hep-ph]}\BibitemShut
  {NoStop}%
%%CITATION = ARXIV:0905.1148;%%
\bibitem [{\citenamefont {De~Simone}\ \emph {et~al.}(2014)\citenamefont
  {De~Simone}, \citenamefont {Giudice},\ and\ \citenamefont
  {Strumia}}]{deSimone:2014pda}%
  \BibitemOpen
  \bibfield  {author} {\bibinfo {author} {\bibfnamefont {A.}~\bibnamefont
  {De~Simone}}, \bibinfo {author} {\bibfnamefont {G.~F.}\ \bibnamefont
  {Giudice}}, \ and\ \bibinfo {author} {\bibfnamefont {A.}~\bibnamefont
  {Strumia}},\ }\href {\doibase 10.1007/JHEP06(2014)081} {\bibfield  {journal}
  {\bibinfo  {journal} {JHEP}\ }\textbf {\bibinfo {volume} {1406}},\ \bibinfo
  {pages} {081} (\bibinfo {year} {2014})},\ \Eprint
  {http://arxiv.org/abs/1402.6287}{arXiv:1402.6287 [hep-ph]}\BibitemShut
  {NoStop}%
%%CITATION = ARXIV:1402.6287;%%
\bibitem [{\citenamefont {Harigaya}\ \emph {et~al.}(2014)\citenamefont
  {Harigaya}, \citenamefont {Kaneta},\ and\ \citenamefont
  {Matsumoto}}]{Harigaya:2014dwa}%
  \BibitemOpen
  \bibfield  {author} {\bibinfo {author} {\bibfnamefont {K.}~\bibnamefont
  {Harigaya}}, \bibinfo {author} {\bibfnamefont {K.}~\bibnamefont {Kaneta}}, \
  and\ \bibinfo {author} {\bibfnamefont {S.}~\bibnamefont {Matsumoto}},\ }\href
  {\doibase 10.1103/PhysRevD.89.115021} {\bibfield  {journal} {\bibinfo
  {journal} {Phys.Rev.}\ }\textbf {\bibinfo {volume} {D89}},\ \bibinfo {pages}
  {115021} (\bibinfo {year} {2014})},\ \Eprint
  {http://arxiv.org/abs/1403.0715}{arXiv:1403.0715 [hep-ph]}\BibitemShut
  {NoStop}%
%%CITATION = ARXIV:1403.0715;%%
\bibitem [{\citenamefont {Ellis}\ \emph {et~al.}(2015)\citenamefont {Ellis},
  \citenamefont {Luo},\ and\ \citenamefont {Olive}}]{Ellis:2015vaa}%
  \BibitemOpen
  \bibfield  {author} {\bibinfo {author} {\bibfnamefont {J.}~\bibnamefont
  {Ellis}}, \bibinfo {author} {\bibfnamefont {F.}~\bibnamefont {Luo}}, \ and\
  \bibinfo {author} {\bibfnamefont {K.~A.}\ \bibnamefont {Olive}},\ }\href@noop
  {} {\  (\bibinfo {year} {2015})},\ \Eprint
  {http://arxiv.org/abs/1503.07142}{arXiv:1503.07142 [hep-ph]}\BibitemShut
  {NoStop}%
%%CITATION = ARXIV:1503.07142;%%
\bibitem [{\citenamefont {Bhattacherjee}\ \emph
  {et~al.}(2014{\natexlab{b}})\citenamefont {Bhattacherjee}, \citenamefont
  {Choudhury}, \citenamefont {Ghosh},\ and\ \citenamefont
  {Poddar}}]{Bhattacherjee:2013wna}%
  \BibitemOpen
  \bibfield  {author} {\bibinfo {author} {\bibfnamefont {B.}~\bibnamefont
  {Bhattacherjee}}, \bibinfo {author} {\bibfnamefont {A.}~\bibnamefont
  {Choudhury}}, \bibinfo {author} {\bibfnamefont {K.}~\bibnamefont {Ghosh}}, \
  and\ \bibinfo {author} {\bibfnamefont {S.}~\bibnamefont {Poddar}},\ }\href
  {\doibase 10.1103/PhysRevD.89.037702} {\bibfield  {journal} {\bibinfo
  {journal} {Phys.Rev.}\ }\textbf {\bibinfo {volume} {D89}},\ \bibinfo {pages}
  {037702} (\bibinfo {year} {2014}{\natexlab{b}})},\ \Eprint
  {http://arxiv.org/abs/1308.1526}{arXiv:1308.1526 [hep-ph]}\BibitemShut
  {NoStop}%
%%CITATION = ARXIV:1308.1526;%%
\bibitem [{\citenamefont {Low}\ and\ \citenamefont {Wang}(2014)}]{Low:2014cba}%
  \BibitemOpen
  \bibfield  {author} {\bibinfo {author} {\bibfnamefont {M.}~\bibnamefont
  {Low}}\ and\ \bibinfo {author} {\bibfnamefont {L.-T.}\ \bibnamefont {Wang}},\
  }\href {\doibase 10.1007/JHEP08(2014)161} {\bibfield  {journal} {\bibinfo
  {journal} {JHEP}\ }\textbf {\bibinfo {volume} {1408}},\ \bibinfo {pages}
  {161} (\bibinfo {year} {2014})},\ \Eprint
  {http://arxiv.org/abs/1404.0682}{arXiv:1404.0682 [hep-ph]}\BibitemShut
  {NoStop}%
%%CITATION = ARXIV:1404.0682;%%
\bibitem [{\citenamefont {Aad}\ \emph {et~al.}(2015{\natexlab{b}})\citenamefont
  {Aad} \emph {et~al.}}]{Aad:2015rba}%
  \BibitemOpen
  \bibfield  {author} {\bibinfo {author} {\bibfnamefont {G.}~\bibnamefont
  {Aad}} \emph {et~al.} (\bibinfo {collaboration} {ATLAS}),\ }\href@noop {} {\
  (\bibinfo {year} {2015}{\natexlab{b}})},\ \Eprint
  {http://arxiv.org/abs/1504.05162}{arXiv:1504.05162 [hep-ex]}\BibitemShut
  {NoStop}%
%%CITATION = ARXIV:1504.05162;%%
\bibitem [{ATL(2015)}]{ATLAS-CONF-2015-013}%
  \BibitemOpen
  \href@noop {} {\emph {\bibinfo {title} {{Search for metastable heavy charged
  particles with large ionisation energy loss in pp collisions at $\sqrt{s}$ =
  8 TeV using the ATLAS experiment}}}},\ \bibinfo {type} {Tech. Rep.}\ \bibinfo
  {number} {ATLAS-CONF-2015-013}\ (\bibinfo {year} {2015})\BibitemShut
  {NoStop}%
%%CITATION = ATLAS-CONF-2015-013 ETC.;%%
\bibitem [{\citenamefont {Farrar}\ and\ \citenamefont
  {Kolb}(1996)}]{Farrar:1995pz}%
  \BibitemOpen
  \bibfield  {author} {\bibinfo {author} {\bibfnamefont {G.~R.}\ \bibnamefont
  {Farrar}}\ and\ \bibinfo {author} {\bibfnamefont {E.~W.}\ \bibnamefont
  {Kolb}},\ }\href {\doibase 10.1103/PhysRevD.53.2990} {\bibfield  {journal}
  {\bibinfo  {journal} {Phys.Rev.}\ }\textbf {\bibinfo {volume} {D53}},\
  \bibinfo {pages} {2990} (\bibinfo {year} {1996})},\ \Eprint
  {http://arxiv.org/abs/astro-ph/9504081}{arXiv:astro-ph/9504081
  [astro-ph]}\BibitemShut {NoStop}%
%%CITATION = ASTRO-PH/9504081;%%
\bibitem [{\citenamefont {Chung}\ \emph {et~al.}(1997)\citenamefont {Chung},
  \citenamefont {Farrar},\ and\ \citenamefont {Kolb}}]{Chung:1997rq}%
  \BibitemOpen
  \bibfield  {author} {\bibinfo {author} {\bibfnamefont {D.~J.}\ \bibnamefont
  {Chung}}, \bibinfo {author} {\bibfnamefont {G.~R.}\ \bibnamefont {Farrar}}, \
  and\ \bibinfo {author} {\bibfnamefont {E.~W.}\ \bibnamefont {Kolb}},\ }\href
  {\doibase 10.1103/PhysRevD.56.6096} {\bibfield  {journal} {\bibinfo
  {journal} {Phys.Rev.}\ }\textbf {\bibinfo {volume} {D56}},\ \bibinfo {pages}
  {6096} (\bibinfo {year} {1997})},\ \Eprint
  {http://arxiv.org/abs/astro-ph/9703145}{arXiv:astro-ph/9703145
  [astro-ph]}\BibitemShut {NoStop}%
%%CITATION = ASTRO-PH/9703145;%%
\bibitem [{\citenamefont {Gambino}\ \emph {et~al.}(2005)\citenamefont
  {Gambino}, \citenamefont {Giudice},\ and\ \citenamefont
  {Slavich}}]{Gambino:2005eh}%
  \BibitemOpen
  \bibfield  {author} {\bibinfo {author} {\bibfnamefont {P.}~\bibnamefont
  {Gambino}}, \bibinfo {author} {\bibfnamefont {G.}~\bibnamefont {Giudice}}, \
  and\ \bibinfo {author} {\bibfnamefont {P.}~\bibnamefont {Slavich}},\ }\href
  {\doibase 10.1016/j.nuclphysb.2005.08.011} {\bibfield  {journal} {\bibinfo
  {journal} {Nucl.Phys.}\ }\textbf {\bibinfo {volume} {B726}},\ \bibinfo
  {pages} {35} (\bibinfo {year} {2005})},\ \Eprint
  {http://arxiv.org/abs/hep-ph/0506214}{arXiv:hep-ph/0506214
  [hep-ph]}\BibitemShut {NoStop}%
%%CITATION = HEP-PH/0506214;%%
\bibitem [{\citenamefont {Sato}\ \emph {et~al.}(2012)\citenamefont {Sato},
  \citenamefont {Shirai},\ and\ \citenamefont {Tobioka}}]{Sato:2012xf}%
  \BibitemOpen
  \bibfield  {author} {\bibinfo {author} {\bibfnamefont {R.}~\bibnamefont
  {Sato}}, \bibinfo {author} {\bibfnamefont {S.}~\bibnamefont {Shirai}}, \ and\
  \bibinfo {author} {\bibfnamefont {K.}~\bibnamefont {Tobioka}},\ }\href
  {\doibase 10.1007/JHEP11(2012)041} {\bibfield  {journal} {\bibinfo  {journal}
  {JHEP}\ }\textbf {\bibinfo {volume} {1211}},\ \bibinfo {pages} {041}
  (\bibinfo {year} {2012})},\ \Eprint
  {http://arxiv.org/abs/1207.3608}{arXiv:1207.3608 [hep-ph]}\BibitemShut
  {NoStop}%
%%CITATION = ARXIV:1207.3608;%%
\bibitem [{\citenamefont {Sato}\ \emph {et~al.}(2013)\citenamefont {Sato},
  \citenamefont {Shirai},\ and\ \citenamefont {Tobioka}}]{Sato:2013bta}%
  \BibitemOpen
  \bibfield  {author} {\bibinfo {author} {\bibfnamefont {R.}~\bibnamefont
  {Sato}}, \bibinfo {author} {\bibfnamefont {S.}~\bibnamefont {Shirai}}, \ and\
  \bibinfo {author} {\bibfnamefont {K.}~\bibnamefont {Tobioka}},\ }\href
  {\doibase 10.1007/JHEP10(2013)157} {\bibfield  {journal} {\bibinfo  {journal}
  {JHEP}\ }\textbf {\bibinfo {volume} {1310}},\ \bibinfo {pages} {157}
  (\bibinfo {year} {2013})},\ \Eprint
  {http://arxiv.org/abs/1307.7144}{arXiv:1307.7144 [hep-ph]}\BibitemShut
  {NoStop}%
%%CITATION = ARXIV:1307.7144;%%
\bibitem [{\citenamefont {Hisano}\ \emph {et~al.}(2004)\citenamefont {Hisano},
  \citenamefont {Matsumoto},\ and\ \citenamefont {Nojiri}}]{Hisano:2003ec}%
  \BibitemOpen
  \bibfield  {author} {\bibinfo {author} {\bibfnamefont {J.}~\bibnamefont
  {Hisano}}, \bibinfo {author} {\bibfnamefont {S.}~\bibnamefont {Matsumoto}}, \
  and\ \bibinfo {author} {\bibfnamefont {M.~M.}\ \bibnamefont {Nojiri}},\
  }\href {\doibase 10.1103/PhysRevLett.92.031303} {\bibfield  {journal}
  {\bibinfo  {journal} {Phys.Rev.Lett.}\ }\textbf {\bibinfo {volume} {92}},\
  \bibinfo {pages} {031303} (\bibinfo {year} {2004})},\ \Eprint
  {http://arxiv.org/abs/hep-ph/0307216}{arXiv:hep-ph/0307216
  [hep-ph]}\BibitemShut {NoStop}%
%%CITATION = HEP-PH/0307216;%%
\bibitem [{\citenamefont {Hisano}\ \emph {et~al.}(2005)\citenamefont {Hisano},
  \citenamefont {Matsumoto}, \citenamefont {Nojiri},\ and\ \citenamefont
  {Saito}}]{Hisano:2004ds}%
  \BibitemOpen
  \bibfield  {author} {\bibinfo {author} {\bibfnamefont {J.}~\bibnamefont
  {Hisano}}, \bibinfo {author} {\bibfnamefont {S.}~\bibnamefont {Matsumoto}},
  \bibinfo {author} {\bibfnamefont {M.~M.}\ \bibnamefont {Nojiri}}, \ and\
  \bibinfo {author} {\bibfnamefont {O.}~\bibnamefont {Saito}},\ }\href
  {\doibase 10.1103/PhysRevD.71.063528} {\bibfield  {journal} {\bibinfo
  {journal} {Phys.Rev.}\ }\textbf {\bibinfo {volume} {D71}},\ \bibinfo {pages}
  {063528} (\bibinfo {year} {2005})},\ \Eprint
  {http://arxiv.org/abs/hep-ph/0412403}{arXiv:hep-ph/0412403
  [hep-ph]}\BibitemShut {NoStop}%
%%CITATION = HEP-PH/0412403;%%
\bibitem [{\citenamefont {Ibarra}\ \emph {et~al.}(2015)\citenamefont {Ibarra},
  \citenamefont {Pierce}, \citenamefont {Shah},\ and\ \citenamefont
  {Vogl}}]{Ibarra:2015nca}%
  \BibitemOpen
  \bibfield  {author} {\bibinfo {author} {\bibfnamefont {A.}~\bibnamefont
  {Ibarra}}, \bibinfo {author} {\bibfnamefont {A.}~\bibnamefont {Pierce}},
  \bibinfo {author} {\bibfnamefont {N.}~\bibnamefont {Shah}}, \ and\ \bibinfo
  {author} {\bibfnamefont {S.}~\bibnamefont {Vogl}},\ }\href {\doibase
  10.1103/PhysRevD.91.095018} {\bibfield  {journal} {\bibinfo  {journal}
  {Phys.Rev.}\ }\textbf {\bibinfo {volume} {D91}},\ \bibinfo {pages} {095018}
  (\bibinfo {year} {2015})},\ \Eprint
  {http://arxiv.org/abs/1501.03164}{arXiv:1501.03164 [hep-ph]}\BibitemShut
  {NoStop}%
%%CITATION = ARXIV:1501.03164;%%
\bibitem [{\citenamefont {Toharia}\ and\ \citenamefont
  {Wells}(2006)}]{Toharia:2005gm}%
  \BibitemOpen
  \bibfield  {author} {\bibinfo {author} {\bibfnamefont {M.}~\bibnamefont
  {Toharia}}\ and\ \bibinfo {author} {\bibfnamefont {J.~D.}\ \bibnamefont
  {Wells}},\ }\href {\doibase 10.1088/1126-6708/2006/02/015} {\bibfield
  {journal} {\bibinfo  {journal} {JHEP}\ }\textbf {\bibinfo {volume} {0602}},\
  \bibinfo {pages} {015} (\bibinfo {year} {2006})},\ \Eprint
  {http://arxiv.org/abs/hep-ph/0503175}{arXiv:hep-ph/0503175
  [hep-ph]}\BibitemShut {NoStop}%
%%CITATION = HEP-PH/0503175;%%
\bibitem [{The ATLAS Collaboration
  ATL-PHYS-PUB-2014-010()}]{ATL-PHYS-PUB-2014-010}%
  \BibitemOpen
  \bibfield  {author} {The ATLAS Collaboration ATL-PHYS-PUB-2014-010,\
  }\href@noop {} {\  (\bibinfo {year} {2014})}\BibitemShut {NoStop}%
\bibitem [{\citenamefont {Liu}\ and\ \citenamefont
  {Tweedie}(2015)}]{Liu:2015bma}%
  \BibitemOpen
  \bibfield  {author} {\bibinfo {author} {\bibfnamefont {Z.}~\bibnamefont
  {Liu}}\ and\ \bibinfo {author} {\bibfnamefont {B.}~\bibnamefont {Tweedie}},\
  }\href {\doibase 10.1007/JHEP06(2015)042} {\bibfield  {journal} {\bibinfo
  {journal} {JHEP}\ }\textbf {\bibinfo {volume} {1506}},\ \bibinfo {pages}
  {042} (\bibinfo {year} {2015})},\ \Eprint
  {http://arxiv.org/abs/1503.05923}{arXiv:1503.05923 [hep-ph]}\BibitemShut
  {NoStop}%
%%CITATION = ARXIV:1503.05923;%%
\bibitem [{\citenamefont {Khachatryan}\ \emph {et~al.}(2015)\citenamefont
  {Khachatryan} \emph {et~al.}}]{CMS:2014wda}%
  \BibitemOpen
  \bibfield  {author} {\bibinfo {author} {\bibfnamefont {V.}~\bibnamefont
  {Khachatryan}} \emph {et~al.} (\bibinfo {collaboration} {CMS}),\ }\href
  {\doibase 10.1103/PhysRevD.91.012007} {\bibfield  {journal} {\bibinfo
  {journal} {Phys.Rev.}\ }\textbf {\bibinfo {volume} {D91}},\ \bibinfo {pages}
  {012007} (\bibinfo {year} {2015})},\ \Eprint
  {http://arxiv.org/abs/1411.6530}{arXiv:1411.6530 [hep-ex]}\BibitemShut
  {NoStop}%
%%CITATION = ARXIV:1411.6530;%%
\bibitem [{CMS(2013)}]{CMS:2013oea}%
  \BibitemOpen
  \href@noop {} {\emph {\bibinfo {title} {{Search for long-lived neutral
  particles decaying to dijets}}}},\ \bibinfo {type} {Tech. Rep.}\ \bibinfo
  {number} {CMS-PAS-EXO-12-038}\ (\bibinfo {year} {2013})\BibitemShut {NoStop}%
%%CITATION = CMS-PAS-EXO-12-038 ETC.;%%
\bibitem [{The(2013)}]{TheATLAScollaboration:2013yia}%
  \BibitemOpen
  \href@noop {} {\emph {\bibinfo {title} {{Search for long-lived, heavy
  particles in final states with a muon and a multi-track displaced vertex in
  proton-proton collisions at $\sqrt{s}$ = 8TeV with the ATLAS detector.}}}},\
  \bibinfo {type} {Tech. Rep.}\ \bibinfo {number} {ATLAS-CONF-2013-092}\
  (\bibinfo {year} {2013})\BibitemShut {NoStop}%
%%CITATION = ATLAS-CONF-2013-092 ETC.;%%
\bibitem [{\citenamefont {Aad}\ \emph {et~al.}(2013{\natexlab{b}})\citenamefont
  {Aad} \emph {et~al.}}]{Aad:2012zx}%
  \BibitemOpen
  \bibfield  {author} {\bibinfo {author} {\bibfnamefont {G.}~\bibnamefont
  {Aad}} \emph {et~al.} (\bibinfo {collaboration} {ATLAS}),\ }\href {\doibase
  10.1016/j.physletb.2013.01.042} {\bibfield  {journal} {\bibinfo  {journal}
  {Phys.Lett.}\ }\textbf {\bibinfo {volume} {B719}},\ \bibinfo {pages} {280}
  (\bibinfo {year} {2013}{\natexlab{b}})},\ \Eprint
  {http://arxiv.org/abs/1210.7451}{arXiv:1210.7451 [hep-ex]}\BibitemShut
  {NoStop}%
%%CITATION = ARXIV:1210.7451;%%
\bibitem [{\citenamefont {Aad}\ \emph {et~al.}(2012{\natexlab{b}})\citenamefont
  {Aad} \emph {et~al.}}]{Aad:2011zb}%
  \BibitemOpen
  \bibfield  {author} {\bibinfo {author} {\bibfnamefont {G.}~\bibnamefont
  {Aad}} \emph {et~al.} (\bibinfo {collaboration} {ATLAS}),\ }\href {\doibase
  10.1016/j.physletb.2011.12.057} {\bibfield  {journal} {\bibinfo  {journal}
  {Phys.Lett.}\ }\textbf {\bibinfo {volume} {B707}},\ \bibinfo {pages} {478}
  (\bibinfo {year} {2012}{\natexlab{b}})},\ \Eprint
  {http://arxiv.org/abs/1109.2242}{arXiv:1109.2242 [hep-ex]}\BibitemShut
  {NoStop}%
%%CITATION = ARXIV:1109.2242;%%
\bibitem [{\citenamefont {Corcella}\ \emph {et~al.}(2001)\citenamefont
  {Corcella}, \citenamefont {Knowles}, \citenamefont {Marchesini},
  \citenamefont {Moretti}, \citenamefont {Odagiri} \emph
  {et~al.}}]{Corcella:2000bw}%
  \BibitemOpen
  \bibfield  {author} {\bibinfo {author} {\bibfnamefont {G.}~\bibnamefont
  {Corcella}}, \bibinfo {author} {\bibfnamefont {I.}~\bibnamefont {Knowles}},
  \bibinfo {author} {\bibfnamefont {G.}~\bibnamefont {Marchesini}}, \bibinfo
  {author} {\bibfnamefont {S.}~\bibnamefont {Moretti}}, \bibinfo {author}
  {\bibfnamefont {K.}~\bibnamefont {Odagiri}},  \emph {et~al.},\ }\href
  {\doibase 10.1088/1126-6708/2001/01/010} {\bibfield  {journal} {\bibinfo
  {journal} {JHEP}\ }\textbf {\bibinfo {volume} {0101}},\ \bibinfo {pages}
  {010} (\bibinfo {year} {2001})},\ \Eprint
  {http://arxiv.org/abs/hep-ph/0011363}{arXiv:hep-ph/0011363
  [hep-ph]}\BibitemShut {NoStop}%
%%CITATION = HEP-PH/0011363;%%
\bibitem [{\citenamefont {Corcella}\ \emph {et~al.}(2002)\citenamefont
  {Corcella}, \citenamefont {Knowles}, \citenamefont {Marchesini},
  \citenamefont {Moretti}, \citenamefont {Odagiri} \emph
  {et~al.}}]{Corcella:2002jc}%
  \BibitemOpen
  \bibfield  {author} {\bibinfo {author} {\bibfnamefont {G.}~\bibnamefont
  {Corcella}}, \bibinfo {author} {\bibfnamefont {I.}~\bibnamefont {Knowles}},
  \bibinfo {author} {\bibfnamefont {G.}~\bibnamefont {Marchesini}}, \bibinfo
  {author} {\bibfnamefont {S.}~\bibnamefont {Moretti}}, \bibinfo {author}
  {\bibfnamefont {K.}~\bibnamefont {Odagiri}},  \emph {et~al.},\ }\href@noop {}
  {\  (\bibinfo {year} {2002})},\ \Eprint
  {http://arxiv.org/abs/hep-ph/0210213}{arXiv:hep-ph/0210213
  [hep-ph]}\BibitemShut {NoStop}%
%%CITATION = HEP-PH/0210213;%%
\bibitem [{\citenamefont {Richter-Was}(2002)}]{RichterWas:2002ch}%
  \BibitemOpen
  \bibfield  {author} {\bibinfo {author} {\bibfnamefont {E.}~\bibnamefont
  {Richter-Was}},\ }\href@noop {} {\  (\bibinfo {year} {2002})},\ \Eprint
  {http://arxiv.org/abs/hep-ph/0207355}{arXiv:hep-ph/0207355
  [hep-ph]}\BibitemShut {NoStop}%
%%CITATION = HEP-PH/0207355;%%
\bibitem [{\citenamefont {Farrar}\ and\ \citenamefont
  {Fayet}(1978)}]{Farrar:1978xj}%
  \BibitemOpen
  \bibfield  {author} {\bibinfo {author} {\bibfnamefont {G.~R.}\ \bibnamefont
  {Farrar}}\ and\ \bibinfo {author} {\bibfnamefont {P.}~\bibnamefont {Fayet}},\
  }\href {\doibase 10.1016/0370-2693(78)90858-4} {\bibfield  {journal}
  {\bibinfo  {journal} {Phys.Lett.}\ }\textbf {\bibinfo {volume} {B76}},\
  \bibinfo {pages} {575} (\bibinfo {year} {1978})}\BibitemShut {NoStop}%
%%CITATION = PHLTA,B76,575;%%
\bibitem [{\citenamefont {Nagata}\ \emph {et~al.}(2015)\citenamefont {Nagata},
  \citenamefont {Otono},\ and\ \citenamefont {Shirai}}]{WIP}%
  \BibitemOpen
  \bibfield  {author} {\bibinfo {author} {\bibfnamefont {N.}~\bibnamefont
  {Nagata}}, \bibinfo {author} {\bibfnamefont {H.}~\bibnamefont {Otono}}, \
  and\ \bibinfo {author} {\bibfnamefont {S.}~\bibnamefont {Shirai}},\
  }\href@noop {} {\  (\bibinfo {year} {2015})},\ \Eprint
  {http://arxiv.org/abs/1506.08206}{arXiv:1506.08206 [hep-ph]}\BibitemShut
  {NoStop}%
%%CITATION = ARXIV:1506.08206;%%
\end{thebibliography}%
%%%%%%%%%%%%%%%%%%%%%%%%%%%%%%%%%%%%%%%%%%%%

\end{document}